\documentclass[11pt]{article}
\usepackage[utf8]{inputenc}
\usepackage{booktabs}
\usepackage[dvipdfmx]{graphicx} \usepackage{bmpsize}
\usepackage{pgfplotstable}
\usepackage{float}
\usepackage{amssymb,amsmath,amsfonts, mathtools}
\usepackage{enumerate}
\usepackage{dsfont}
\usepackage[sort,compress]{cite}
\usepackage{verbatim}
\usepackage{amsfonts,euscript,amssymb,stmaryrd,braket}
\usepackage{graphics,tikz}
\usetikzlibrary{arrows,decorations.markings,patterns}
\usepackage{caption}
\usepackage{subcaption}
\usepackage{slashed}
\definecolor{hyperref}{RGB}{026,028,185}
\usepackage[bookmarks=true,colorlinks=true,linkcolor=hyperref,citecolor=hyperref,urlcolor=hyperref,bookmarksnumbered]{hyperref}
 \usepackage{booktabs}
 \usepackage{multirow}
\usepackage{tabularx}
\usepackage{longtable}
\usepackage{bbold,bbding}
\usepackage{amsthm} 
\usepackage{esint}

\newcommand{\bal}{\begin{equation}\begin{aligned}}
\newcommand{\eal}{\end{aligned} \end{equation}}

 \def\clock{{\count0=\time
           \divide\count0 60
           \ifnum\count0<10 0\fi\the\count0
           \multiply\count0 -60 \advance\count0 \time
           :\ifnum\count0<10 0\fi \the\count0
         }}
\newcommand{\timestamp}{{\small\vbox{\hbox{\tt\jobname.tex}
\hbox{\the\day/\the\month/\the\year, \clock}}}}

\newcommand{\ba}{\begin{eqnarray}}
\newcommand{\ea}{\end{eqnarray}}

\newcommand{\be}{\begin{equation}}
\newcommand{\ee}{\end{equation}}

\makeatletter
\let\old@startsection=\@startsection
\let\oldl@section=\l@section
\renewcommand{\@startsection}[6]{\old@startsection{#1}{#2}{#3}{#4}{#5}{#6\mathversion{bold}}}
\renewcommand{\l@section}[2]{\oldl@section{\mathversion{bold}#1}{#2}}
\makeatother

\numberwithin{equation}{section}
\newcommand{\Pf}{\textup{Pf}}

\usepackage{color}
\newcommand{\colb}[1]{\textcolor{blue}{#1}}

\def\no{\nonumber}

\def \adss {$AdS_5 \times S^5$\ }

\newcommand{\beq}{\begin{equation}}
\newcommand{\eeq}{\end{equation}}

\renewcommand{\d}{\delta}

\newcommand{\e}{\epsilon}

\renewcommand{\S}{\Sigma}

\newcommand{\Tr}{\textup{Tr}}

\begin{document}
\renewcommand{\thefootnote}{\arabic{footnote}}

\overfullrule=0pt
\parskip=2pt
\parindent=12pt
\headheight=0in \headsep=0in \topmargin=0in \oddsidemargin=0in

\vspace{ -3cm} \thispagestyle{empty} \vspace{-1cm}
\begin{flushright} 
\footnotesize
{HU-EP-19/29}
\\
{Imperial-TP-EV-2019-04}
\end{flushright}%

\begin{center}
\vspace{1.2cm}
{\Large\bf \mathversion{bold}
{New linearization and reweighting for simulations of string sigma-model on the lattice}
}
 
\author{ABC\thanks{XYZ} \and DEF\thanks{UVW} \and GHI\thanks{XYZ}}
 \vspace{0.8cm} {
 L.~Bianchi$^{a}$\footnote{ {\tt  lorenzo.bianchi@qmul.ac.uk}}, V.~Forini$^{b,c}$\footnote{{\tt valentina.forini@city.ac.uk}}, B.~Leder$^{c,}$\footnote{{\tt $\{$leder,philipp.toepfer$\}$@\,physik.hu-berlin.de}}, P.~T\"opfer$^{c,3}$, E.~Vescovi$^{e}$\footnote{ {\tt e.vescovi@imperial.ac.uk}}}
 \vskip  0.5cm

\small
{\em
%$^{a}$  II. Institut f\"ur Theoretische Physik, Universit\"at Hamburg,\\ Luruper Chaussee 149, 22761 Hamburg, Germany   
%\vskip 0.05cm
$^{a}$  Centre for Research in String Theory, %School of Physics and Astronomy,\\
Queen Mary University of London\\ Mile End Road, London E1 4NS, United Kingdom
\vskip 0.05cm
$^{b}$ Department of Mathematics, City, University of London,\\
Northampton Square, EC1V 0HB London, United Kingdom
\vskip 0.05cm
  $^{c}$  
Institut f\"ur Physik, Humboldt-Universit\"at zu Berlin and IRIS Adlershof, \\Zum Gro\ss en Windkanal 6, 12489 Berlin, Germany  
\vskip 0.05cm
%$^{e}$ Institute of Physics, University of S\~{a}o Paulo 05314-970, S\~{a}o Paulo, Brazil \vskip 0.05cm
$^{e}$ The Blackett Laboratory, Imperial College, London SW7 2AZ, United Kingdom
}
\normalsize

\end{center}

\vspace{0.3cm}
\begin{abstract}  
We  study
% RESUB , first presented in~\cite{Forini:2017mpu},
 the discretized worldsheet of Type IIB strings in the Gubser-Klebanov-Polyakov background in a new setup, which eliminates a complex phase previously detected in the fermionic determinant. A sign ambiguity remains, which a study of the fermionic spectrum shows to be related to Yukawa-like terms, including those present in the original Lagrangian before the linearization standard in a lattice QFT approach.  Monte Carlo simulations are performed in a large  region of the parameter space, where the sign problem starts becoming severe and instabilities appear due to the zero eigenvalues of the fermionic operator.  To face these problems, simulations are conducted using the absolute value of a fermionic Pfaffian obtained introducing a small twisted-mass term, acting as an infrared regulator,  into the action.
%modified with an infrared regulator. 
The sign of the Pfaffian and the low modes of the quadratic fermionic operator are then taken into account by a reweighting procedure of which we discuss the impact on the measurement of the observables. In this setup we study bosonic and fermionic correlators and observe a divergence in the latter, which we argue - also via a one-loop analysis in lattice perturbation theory -  to originate from the U(1)-breaking of our Wilson-like discretization for the fermionic sector.

\end{abstract}

\newpage

%%%%%%%%%%%%%%%%%%%%%%%%%%%%%%%%%%%%%%%%%%%%%
%%%%%%%%%%%%%%%%%%%%%%%%%%%%%%%%%%%%%%%%%%%%%
\tableofcontents
  
\section{Introduction and Discussion}

Lattice field theory methods are already employed for some time in the broad context of AdS/CFT (see e.g.~\cite{Catterall_physrept, Schaich:2015ppr,Joseph:2015xwa, Bergner:2016sbv, Schaich:2016jus, Berkowitz:2016jlq,Rinaldi:2017mjl, Catterall:2017lub, Schaich:2018mmv}), and more recently also  from the point of view of string sigma-models in AdS backgrounds~\cite{Roiban, POS2015,Bianchi:2016cyv, Forini:2017mpu, Forini:2017ene}. 
In this case the focus has been on a particularly central model for the AdS/CFT community, the  string worldsheet dual to a light-like cusped Wilson loop. The renormalization of the latter is governed by the cusp anomalous dimension, an observable of crucial importance in all gauge theories and also in the maximally supersymmetric one, $\mathcal{N}=4$ super Yang-Mills in four dimensions. Its non-perturbative behavior is there accessible exactly, when using the assumption and the tools of integrability~\cite{BES,Basso:2013vsa,Basso:2013aha,Fioravanti:2015dma, Bonini:2018mkg}. 
From the perspective of superstring theory, the relevant sigma-model - a Green-Schwarz  action in $AdS_5\times S^5$ background with Ramond-Ramond flux - is a complicated, highly non-linear two-dimensional field theory which is not known how to solve exactly  and has been approached perturbatively, so far up to two-loop level,  in a semiclassical way. Applying lattice field theory methods for its non-perturbative investigation appears to be a formidable benchmark test 
for a wider program which aims at using this approach to numerical holography in much more general cases, for which exact predictions do not exist. 
This is particularly true since, as from the preliminary results of Ref.~\cite{Bianchi:2016cyv}, this model appears to present in a single setup many of the challenges of lattice investigations in QFT, such as e.g. symmetry-breaking discretizations, numerical instabilities and even a complex phase problem.  In this paper we make significant steps in addressing these points. 

The model under study is the AdS-lightcone gauge-fixed, Type IIB Green-Schwarz  superstring action~\cite{MT2000,MTT2000} describing fluctuations about the  Gubser-Klebanov-Polyakov background~\cite{Gubser:2002tv}, and was worked out explicitly in~\cite{Giombi}. From the point of view of an investigation with lattice field theory methods, it is a non-linear action with no gauge degrees of freedom and where fermions, which couple via a quartic interaction, do not carry (Lorentz) spinor indices but are just a set of anticommuting scalars. A global $SO(6)\times SO(2)$ symmetry is explicitly realized.  In continuum perturbation theory, results are available up to  two loop order~\cite{Giombi,Giombi:2010bj} (see also~\cite{Giombi:2010zi}). 

The analysis of Refs.~\cite{POS2015,Bianchi:2016cyv} presented a discretization of the (linearized) model based on a Wilson-like treatment of the fermionic sector which was tested via a one-loop analysis in lattice perturbation theory. An estimation of the (derivative) of the cusp anomaly of $\mathcal{N}=4$ super Yang-Mills  was provided, via a measurement of the vacuum expectation value of the relevant action  in terms of simulations performed employing a Rational Hybrid Monte Carlo (RHMC) algorithm. %~\colr{[you may ignore and set a full stop], as well as of the  mass of two AdS excitations transverse to the relevant null cusp classical string solution.} 
In this context, the (dimensionless) coupling constant is the effective string tension $g=\frac{R^2}{4\pi\alpha'}\equiv\frac{\sqrt{\lambda}}{4\pi}$, where $R$ is the common radius of $AdS_5$ and $S^5$ and  $\lambda$ is the 't~Hooft coupling,  and the perturbative expansion is a series in inverse powers of the effective string tension. Therefore, the string sigma-model is weakly coupled for large values of $g$ and in this regime, a good qualitative agreement was observed with the exact predictions obtained via integrability methods.  In the case of higher-order fermionic interactions, one proceeds first linearizing the model via the introduction a set  of auxiliary fields,  then integrates out the fermionic determinant/Pfaffian re-exponentiating it in terms of a set of bosonic fields called pseudio-fermions and letting it become part of the Boltzmann weight of configurations in the statistical ensemble.
It was observed in~\cite{Bianchi:2016cyv}  that the nature of the quartic interaction -- in which a ``repulsive'' potential appears -- is responsible for the appearance of a non-hermitian piece in the linearized Lagrangian, which eventually gives rise to a \emph{complex phase} in the fermionic Pfaffian. For lower values of $g$, namely when the string sigma-model is strongly coupled, a severe sign problems appears. 

In what follows we discuss a new linearization of the four-fermion term~\colb{\footnote{This new linearization has been presented at various conferences and in the proceedings~\cite{Forini:2017mpu}.}} which eliminates the complex phase -- albeit not the sign problem (the latter is expected in most systems with interacting fermions).  We will proceed  % \colr{is inspired by~\cite{Catterall:2015zua} (see also~\cite{Catterall:2016dzf})} and it 
via an algebraic manipulation of the original fermionic Lagrangian. The resulting quadratic fermionic operator $O_F$ is antisymmetric and ``$\gamma_5$-hermitian'', two properties which ensure a real, non-negative $\det O_F$ and a \emph{real} Pfaffian $({\rm Pf}\,O_F)^2=\det O_F \geq0$. This is quite crucial, as eliminating the complex phase allows to eliminate a systematic error in measurements, in particular in the so-called reweighting procedure (see Section~\ref{sec:simulations_strong} below), in which the possibly present phase would have to be calculated explicitly~\footnote{An efficient evaluation of complex determinants for arbitrarily large matrices is highly non trivial. For this reason, in~\cite{Bianchi:2016cyv} this has been done only for small lattices. It was there observed that the reweighting had no effect on the central value of the observables  under study, therefore the phase was omitted from the simulations when taking the continuum limit ($N\to\infty$). In absence of data for  larger lattices the possible systematic error related to this procedure was not assessed.}.  Because of the sign ambiguity in ${\rm Pf} \,O_F=\pm\sqrt{\det O_F}$, a sign problem may still remain, which is in fact the case. 
Below - via a study of the fermionic spectrum~\cite{Forini:2017mpu} - we show that the sign ambiguity appears to be related to the Yukawa-like terms, including those present before linearization, and therefore in the original Lagrangian. By looking at the lowest eigenvalue for the squared fermionic operator $\hat  O_F^\dagger \hat O_F$  in a large region of the parameter space, we also observe below that sign flips are extremely unlikely in an  interesting regime of the coupling, $g\simeq10$.  

Together with the sign problem, for lower values of $g$ the zero eigenvalues of the fermionic operator cause numerical instabilities, due to the non-convergence of the inverter for the fermionic matrix.  Mimicking  the twisted-mass reweighting procedure of~\cite{Luscher:2008tw} we perform simulations using the absolute value of a fermionic Pfaffian modified with an infrared regulator.  The sign of the Pfaffian and the low modes of $O_F$ are then taken into account by a reweighting procedure of which we discuss in details the impact on the measurement of the observables. We are confident that simulations of the model in this setup are stable in a very large region of the parameter space $g\geq2$, with in principle no obvious obstacle for simulations at even smaller value of $g$.   The sign problem becomes severe for $g<5$, which makes measurements unreliable in this region. However,  it is very interesting to observe that the sign-reweighting seems \emph{not} to have effect on the measured observables, and it would be important to investigate why this happens further. 

Below we  investigate two kinds of observables --- bosonic and fermionic correlators of the field excitations about the Gubser-Klebanov-Polyakov background~\cite{Gubser:2002tv} -- and observe a linear divergence in the measurements of the fermionic masses. This is reminiscent of  a typical phenomenon occurring in lattice QCD for quark masses in the case of Wilson fermions, an additive renormalization which manifests itself as a power (linear) divergence in the lattice spacing and it is related to the fact that the lattice action for fermions breaks chiral symmetry (see e.g.~\cite{montvay}). In our case, it is natural to trace back the observed divergence to the fact that our discretization breaks the $U(1)$ part of the original $SO(6)\times U(1)$ symmetry of our model. We argue this in details below,  using numerics and the relation to the bosonic counterpart of this divergence -- the linearly divergent one-point functions of the two AdS excitations transverse to the relevant null cusp classical string solution. The latter are calculated at leading order in lattice perturbation theory in Appendix~\ref{app:onepoint}. 
 
An immediate and crucial outlook of the analysis here presented is the necessity of a redefinition of the continuum limit, which should take into account the infinite mass renormalization observed and therefore a possible tuning of the ``bare'' mass parameter of the theory (the light-cone momentum $P_+$, which we redefine as $m$ below). One way to proceed is by studying the violation of the continuum Ward identities  on the lattice and explicitly checking that these violations vanish in the continuum limit. It would be also mostly interesting to investigate discretizations of the fermionic action (e.g. inspired to Ginsparg-Wilson fermions) which may preserve a larger symmetry group on the lattice~\footnote{We thank Agostino Patella for discussions on this.}.  % \colb{(if it is just extrapolation)}.

This paper proceeds with a presentation of the details on the algebraic manipulation of the Lagrangian and its novel linearization (Section~\ref{sec:linearization}), an analysis of the spectrum of the fermionic operator (Section~\ref{sec:spectrum}), a study of bosonic and fermionic correlators (Section~\ref{sec:simulations_strong}) and an analysis of the impact of reweighting procedure on the observables (Section~\ref{sec:reweighting}). Appendices collect notation and useful details for deriving the  fermionic linearization (Appendix~\ref{app:continuum}) as well as the evaluation at leading order in lattice perturbation theory of the non-trivial one-point function $\langle x\rangle$   (Appendix \ref{app:onepoint}).

\section{Linearization and phase-free Pfaffian}
\label{sec:linearization}

The Euclidean superstring action  in AdS-lightcone gauge-fixing~\cite{MT2000,MTT2000}  describing quantum  fluctuations around  the null-cusp background  in \adss reads~\cite{Giombi}
\begin{eqnarray}\nonumber
&& \!\!\!\!\!\!\!\!\!\!\!\!
S_{\rm cusp}=g \int dt ds~ \Big\{ 
%\mathcal{L}_{\rm cusp}\\\nonumber
%&& \!\!\!\!\!\! 
%\mathcal{L}_{\rm cusp} = 
|\partial_{t}x+\textstyle{\frac{1}{2}}x|^{2}+\frac{1}{ {z}^{4}} |\partial_{s} {x}-\textstyle{\frac{1}{2}} {x}|^{2}+\left(\partial_{t}z^{M}+\frac{1}{2} {z}^{M}+\frac{i}{ {z}^{2}} {z}_{N} {\eta}_{i}\left(\rho^{MN}\right)_{\phantom{i}j}^{i} {\eta}^{j}\right)^{2}\\\label{S_cusp}
 && \!\!\!\!\!\!\!\!\!\!\!\!
+\frac{1}{ {z}^{4}}\left(\partial_{s} {z}^{M}-\textstyle{\frac{1}{2}} {z}^{M}\right)^{2}  %\\
  +i\left( {\theta}^{i}\partial_{t}{\theta}_{i}+ {\eta}^{i}\partial_{t}{\eta}_{i}+ {\theta}_{i}\partial_{t}{\theta}^{i}+ {\eta}_{i}\partial_{t} {\eta}^{i}\right)-\textstyle{\frac{1}{{z}^{2}}}\left( {\eta}^{i}{\eta}_{i}\right)^{2}  \\\nonumber
 &&  \!\!\!\!\!\!\!\!\!\!\!\!
 +2i\Big[\textstyle{\frac{1}{z^{3}}}z^{M} {\eta}^{i}\left(\rho^{M}\right)_{ij}
 \left(\partial_{s} \theta^j-\textstyle{\frac{1}{2}} \theta^j-\frac{i}{{z}} {\eta}^{j}\left(\partial_{s} {x}-\frac{1}{2} {x}\right)\right)
 %\\
% &&  \!\!\!\!\!\!
\textstyle{+\frac{1}{{z}^{3}}{z}^{M}{\eta}_{i} (\rho_{M}^{\dagger} )^{ij}\left(\partial_{s}{\theta}_{j}-\frac{1}{2}{\theta}_{j}+\frac{i}{{z}}{\eta}_{j}\left(\partial_{s}{x}-\frac{1}{2}{x}\right)^{*}\right)\Big]\,\Big\}}\, \label{cuspaction}
\end{eqnarray}
where $x,x^*$ are two bosonic fields transverse  to the subspace $AdS_3$  of the classical solution and $z^M\, (M=1,\cdots, 6)$, with  $z=\sqrt{z_M z^M}$, are the six cartesian coordinates of the sphere $S^5$.   %As mentioned above,  the Lagrangian above neither contains gauge fields nor actual fermions. Indeed, 
 The Gra\ss mann-odd fields $\theta_i,\eta_i,\, i=1,2,3,4$ are complex variables (no Lorentz spinor indices appear) such that $\theta^i = (\theta_i)^\dagger,$ $\eta^i = (\eta_i)^\dagger$,  transforming in the fundamental representation of the $SU(4)$ R-symmetry group. The matrices $\rho^{M}_{ij} $ are the off-diagonal
blocks of $SO(6)$ Dirac matrices $\gamma^M$ in  chiral representation, 
and
$(\rho^{MN})_i^{\hphantom{i} j} = (\rho^{[M} \rho^{\dagger N]})_i^{\hphantom{i} j}$ are  the
$SO(6)$ generators. 
Under the $U(1)$ symmetry, the fields $z^M$ are neutral , $\theta^i$ and $\eta^i$ have opposite
charges and the charge of $\eta_i$ ($\eta^i$) is half the charge of $x$ ($x^*$). 
In the action \eqref{S_cusp} a massive parameter ($\sim P_+$) is missing, which we restore below in \eqref{Scuspquadratic} defining it as~$m$. 
% and introducing its dimensionless counterpart $M=a\,m$, where $a$ is the lattice spacing.  
%We emphasize that, in \eqref{S_cusp},  local bosonic (diffeomorphism) and fermionic ($\kappa$-) symmetries originally present have been fixed.}
 
%
As standard, to take into account the fermionic contribution in the case of higher-order interactions one first linearizes the corresponding Lagrangian, making it quadratic in fermions,  and then formally integrates out  the Gra\ss mann-odd fields letting their determinant - here, a Pfaffian -  to enter  the Boltzmann weight of each configuration through re-exponentiation
\begin{equation}\label{fermionsintegration}
\!\!\!\! \int \!\! D\Psi~ e^{-\textstyle\int dt ds \,\Psi^T O_F \Psi}={\rm Pf}\,O_F~~\longrightarrow ~~(\det O_F\,O^\dagger_F)^{\frac{1}{4}}= \int \!\!D\xi D\bar\xi\,e^{-\int dt ds\, \bar\xi(O_FO^\dagger_F)^{-\frac{1}{4}}\,\xi}~,
 \end{equation}
 where the replacement is needed in the case of non-positive-definite Pfaffian. 
 
% the straightforward way to linearize the quartic fermionic interactions in~\eqref{S_cusp} is to introduce  a set of $7$ real auxiliary  fields, one scalar $\phi$ and a  $SO(6)$ vector field $\phi_M$ as in~\colb{ROIBAN,CITEUS}
%\begin{eqnarray}\label{HubbardStratonovich}
%&& \!\!\!\!\!\!\!
%\exp \Big\{-g\int dt ds  \Big[-\textstyle{\frac{1}{{z}^{2}}}\left( {\eta}^{i}{\eta}_{i}\right)^{2}  +\Big(\textstyle{\frac{i}{ {z}^{2}}} {z}_{N} {\eta}_{i}{\rho^{MN}}_{\phantom{i}j}^{i} {\eta}^{j}\Big)^{2}\Big]\}\\\nonumber
%&& 
%\sim\,\int D\phi D\phi^M\,\exp\Big\{-  g\int dt ds\,[\textstyle\frac{1}{2}{\phi}^2+\frac{\sqrt{2}}{z}\phi\,\eta^2 +\frac{1}{2}({\phi}_M)^2-i\,\frac{\sqrt{2}}{z^2}\phi^M {z}_{N} \,\big(i \,{\eta}_{i}{\rho^{MN}}_{\phantom{i}j}^{i} {\eta}^{j}\big)]\Big\}~.
%\end{eqnarray}
%The second quartic interaction above squares an hermitian bilinear, $\Big(i\,\eta_i {\rho^{MN}}^i{}_j \eta^j\Big)^\dagger=i\eta_j\,{\rho^{MN}}^j{}_i\,\eta^i$, and comes in the exponential as a ?repulsive? potential. This has the final effect of an imaginary part in the auxiliary
%
% due to $\Big(i\,\eta_i {\rho^{MN}}^i{}_j \eta^j\Big)^\dagger=i\eta_j\,{\rho^{MN}}^j{}_i\,\eta^i$, the corresponding Yukawa term makes the linearized Lagrangian not hermitian. A complex phase is then easily detected in the resulting Pfaffian , and the the  only  question being whether the latter is treatable via standard reweighing. Below we find evidence that %, at least in this setting, 
%this is not the case.
To linearize, we focus on the part of the Lagrangian in \eqref{S_cusp} which is quartic in fermions
 \begin{equation}\label{eq:quarticaction}
  \mathcal{L}_4=\frac{1}{z^2}\left[- (\eta^2)^2+\left(i\, \eta_i {(\rho^{MN})^i}_j n^N \eta^j\right)^2\right]\,\,,
\end{equation}
where $n^M=\frac{z^M}{z}$. Notice the plus sign in front of the second term in \eqref{eq:quarticaction}, which squares an hermitian bilinear $(i\,\eta_i {\rho^{MN}}^i{}_j \eta^j)^\dagger=i\eta_j\,{\rho^{MN}}^j{}_i\,\eta^i$~\cite{Bianchi:2016cyv}. Then the standard Hubbard-Stratonovich transformation
\begin{eqnarray}\label{HubbardStratonovich}
&& \!\!\!\!\!\!\!
\exp \Big\{-g\int dt ds  \Big[-\textstyle{\frac{1}{{z}^{2}}}\left( {\eta}^{i}{\eta}_{i}\right)^{2}  +\Big(\textstyle{\frac{i}{ {z}^{2}}} {z}_{N} {\eta}_{i}{\rho^{MN}}_{\phantom{i}j}^{i} {\eta}^{j}\Big)^{2}\Big]\}\\\nonumber
&& 
\sim\,\int D\phi D\phi^M\,\exp\Big\{-  g\int dt ds\,[\textstyle\frac{1}{2}{\phi}^2+\frac{\sqrt{2}}{z}\phi\,\eta^2 +\frac{1}{2}({\phi}_M)^2-i\,\frac{\sqrt{2}}{z^2}\phi^M {z}_{N} \,\big(i \,{\eta}_{i}{\rho^{MN}}_{\phantom{i}j}^{i} {\eta}^{j}\big)]\Big\}~
\end{eqnarray}
generates a non-hermitian term, the last one above, resulting in a  complex-valued Pfaffian for the fermionic operator. Here we provide a solution to this problem, obtaining a \emph{real-valued} Pfaffian via an alternative procedure, where the first step  is  rewriting the Lagrangian \eqref{eq:quarticaction} with a  procedure inspired by~\cite{Catterall:2015zua}. There,  a simpler  action with $SO(4)$ four-fermion terms  in three dimensions was considered (see also the four-dimensional $SU(4)$ counterpart in~\cite{Catterall:2016dzf}). Our Lagrangian \eqref{eq:quarticaction} is invariant under $SU(4)\times U(1)$ transformations and this requires a generalization of \cite{Catterall:2015zua} that preserves this symmetry. Let us start by eliminating the matrices $\rho^{MN}$ from the second term of \eqref{eq:quarticaction} in favour of $\rho^M$, which after some $\rho$-matrices manipulations leads to
\begin{align}\label{newquarticaction}
\mathcal{L}_4=\frac{1}{z^2}\left(- 4\, (\eta^2)^2+2\left|\eta_i (\rho^N)^{ik} n_N \eta_k\right|^2\right)\,.
\end{align}
%where the plus sign in front of the second term still prevents a real Pfaffian after the Hubbard-Stratonovich transformation. 
We then define a duality transformation, reminiscent of the standard Hodge duality but adapted to our particular case. Given ${\S_i}^j\equiv\eta_i \eta^j$ the dual matrix $\tilde{\S}_j{}^i$ is defined by
\begin{align}\label{sigmatilde}
\tilde{\S}_j{}^i=n_N n_L(\rho^N)^{ik}(\rho^L)_{jl} {\S_k}^l\,\,.
\end{align}
Notice that $\tilde{\tilde \S}=\S$ and ${\S^i}_j\equiv ({\S_i}^j)^\dagger={\S_j}^i $. One can then easily rewrite \eqref{newquarticaction} as 
\begin{align}\label{lagsig}
 \mathcal{L}_4=\frac{2}{z^2}\Tr\left( \S\S+ \tilde \S\tilde \S- \S\tilde \S\right)\,\,,
\end{align}
where the trace is over $SU(4)$ fundamental indices. Although we split the first two terms in \eqref{lagsig} to exhibit the neutrality of the Lagrangian under duality transformation, it is useful to keep in mind that $\Tr \tilde \S\tilde \S=\Tr \S\S$. Since we want to write down a Lagrangian as the sum of two terms squared, it is natural to introduce the self- and antiself-dual part of $\S$ 
\begin{align}
 {\S_{\pm}}=\S \pm \tilde \S
\end{align}
such that $\tilde\S_{\pm}=\pm \S_{\pm}$. Now the crucial, though elementary fact that $\Tr \S_{\pm}\S_{\pm}=2\Tr \left(\S\S\pm  \S\tilde \S\right)$ gives us some freedom in the choice of the sign in the Lagrangian~\footnote{It is worth emphasizing that there is neither ambiguity nor arbitrariness in the double sign present in \eqref{lagsig}: Writing the Lagrangian in terms of the self-dual part of $\S$ requires the minus sign, writing it in terms of the antiself-dual part requires the plus sign.}, since
\begin{align}\label{lagpm}
 \mathcal{L}_4=\frac{1}{z^2}\Tr\left(4 \S\S\mp \S_{\pm} \S_{\pm} \pm 2 \S\S \right)~.
\end{align}
This last equation proves that the complex phase is an artefact of our naive linearization. Indeed, \eqref{lagpm} provides two equivalent forms of the same action, one which would lead to a phase problem and one which would not. Choosing the latter, i.e. the one involving $\S_+$, we obtain for the quartic Lagrangian the expression
\begin{equation}
 \mathcal{L}_4=\frac{1}{z^2}\left(- 6\, (\eta^2)^2 - {\S_{+}}_i^j{\S_{+}}_j^i \right)~.
\end{equation}
In this form the Lagrangian is suitable for the following Hubbard-Stratonovich transformation
%
 %. In particular we have
 \bal\label{HubbardStratonovich}
&\exp \Big\{-g\int dt ds  \Big[-\textstyle{\frac{1}{z^2}\left(- 6\, (\eta^2)^2 - {\S_{+}}_i^j{\S_{+}}_j^i \right)}\Big]\Big\}\\ 
&\sim ~\int D\phi D\phi^M\,\exp\Big\{-  g\int dt ds\,[\textstyle \frac{12}{z} \eta^2 \phi +6\phi^2+\frac{2}{z} {\S_+}^i_j \phi^j_i +\phi^i_j \phi^j_i ]\Big\}~,
\eal
where $\phi$ is real and $\phi^i_j$ can be thought of as a $4\times 4$ complex hermitian matrix with 16 real degrees of freedom~\footnote{The proof of \eqref{HubbardStratonovich} is based on these properties, the split of ${\S_{+}}_i^j$ and $\phi^i_j$ with $i\neq j$ into real and imaginary parts and the Gaussian integration formula over real variables.}. Therefore the new linearization proposed here introduces a total of 17 auxiliary fields.

The final form of the Lagrangian is then
\bal\label{Scuspquadratic}
{\cal L} &=  {| \partial_t {x} + {\frac{m}{2}}{x} |}^2 + \frac{1}{{ z}^4}{\big| \partial_s {x} -\frac{m}{2}{x} |}^2
+ (\partial_t {z}^M + \frac{m}{2}{z}^M )^2 + \frac{1}{{ z}^4} (\partial_s {z}^M -\frac{m}{2}{z}^M)^2
\\
&+6\phi^2+\phi^i_j \phi^j_i+\psi^T O_F \psi\
\eal
with $\psi\equiv\left(\theta^{i},\theta_{i},\eta^{i},\eta_{i}\right)$ and
%%%%%%%%%%%%%%%%%%%%%%%%%%%%%%%%%%%%%%%%%%%%%
\begin{equation} \label{OF}
O_F =\left(\begin{array}{cccc}
0 & \mathrm{i}\partial_{t} & -\mathrm{i}\rho^{M}\left(\partial_{s}+\frac{m}{2}\right)\frac{{z}^{M}}{{z}^{3}} & 0\\
\mathrm{i}\partial_{t} & 0 & 0 & -\mathrm{i}\rho_{M}^{\dagger}\left(\partial_{s}+\frac{m}{2}\right)\frac{{z}^{M}}{{z}^{3}}\\
\mathrm{i}\frac{{z}^{M}}{{z}^{3}}\rho^{M}\left(\partial_{s}-\frac{m}{2}\right) & 0 & 2\frac{{z}^{M}}{{z}^{4}}\rho^{M}\left(\partial_{s}{x}-m\frac{{x}}{2}\right) & \mathrm{i}\partial_{t}-A^{T}\\
0 & \mathrm{i}\frac{{z}^{M}}{{z}^{3}}\rho_{M}^{\dagger}\left(\partial_{s}-\frac{m}{2}\right) &\mathrm{i}\partial_{t}+A & -2\frac{{z}^{M}}{{z}^{4}}\rho_{M}^{\dagger}\left(\partial_{s}{x}^\ast-m\frac{{x}}{2}^\ast\right)
\end{array}\right)~,
\end{equation}
where
\begin{align}\label{A}
A&=-\frac{6}{z}\phi + \frac{1}{z}\tilde{\phi}+\frac{1}{z^{3}}\rho^\ast_{N}\tilde{\phi}^{T}\rho^{L}z^{N}z^{L}+\mathrm{i}\frac{z^{N}}{z^2}\rho^{MN}\partial_{t}z^{M}, \qquad\qquad
\tilde{\phi}&\equiv \left(\tilde{\phi}_{ij}\right)\equiv \left(\phi^{i}_{j}\right).
\end{align}

The discretization that we will adopt here was presented in \cite{Bianchi:2016cyv}. There, it was observed that it is a priori not possible to remove fermion doublers while maintaining all the symmetries of the model and  preventing complex phases to appear in the determinant. A ``minimal-breaking'' solution preserves the $SU(4)$ global symmetry of the Lagrangian and \emph{breaks} the $U(1)$~\footnote{Another possible discretization, also used in~\cite{Bianchi:2016cyv}, breaks both $SO(6)$ and $U(1)$ symmetries.}, and it consists in adding a Wilson-like term in the main diagonal of the fermionic operator. In lattice perturbation theory, this discretization reproduces in the continuum limit $a\to 0$ the large $g$, one-loop value of the cusp anomalous dimension~\cite{Bianchi:2016cyv}. As the new linearization affects off-diagonal terms ($A$-terms), we can simply proceed with the proposal in~\cite{Bianchi:2016cyv} for the discretized fermionic operator
\begin{eqnarray}\nonumber
\!\!\!\!\!\!\!\!\!\!\!\!\!\!\!\!
{\hat O_F}&\!\!=\!\!&\left(\begin{array}{cccc}
    \!\!\!\!  W_+      & \!\!\!\! - \mathring{p_0} \mathbb{1}      &\!\!\!\! (\mathring{p_1}-i\frac{m}{2} )\rho^M \frac{z^M}{z^3}  & \!\!\!\! 0 \\
 \!\!\!\!     - \mathring{p_0}\mathbb{1}  & \!\! -W_+^{\dagger}          &\!\!\!\! 0        & \!\!\!\! \rho_M^\dagger(\mathring{p_1}-i\frac{ m}{2})\frac{z^M}{z^3} \\
    \!\!\!\!  -( \mathring{p_1}+i\frac{m}{2} )\rho^M \frac{z^M}{z^3}   &\!\!\!\! 0          &\!\!\!\! 2\frac{{z}^{M}}{{z}^{4}}\rho^{M}\left(\partial_{s}{x}-m\frac{{x}}{2}\right) +W_-     &\!\!\!\! - \mathring{p_0} \mathbb{1}-A^T\\
   \!\!\!\!   0      &\!\!\!\! -\rho_M^\dagger( \mathring{p_1}+i\,\frac{m}{2})\frac{z^M}{z^3}   &\!\!\!\! - \mathring{p_0}\mathbb{1}+A    &\!\!\!\! -2\frac{{z}^{M}}{{z}^{4}}\rho_{M}^{\dagger}\left(\partial_{s}{x}^*-m\frac{{x}}{2}^*\right) -W_-^\dagger
          \end{array}\right)\\
          \label{OFgen}
          \end{eqnarray}
       with~\cite{montvay}  
%where $M=\frac12 a m$, 
\be\label{pcircphat}
 \mathring{p}_\mu=\frac{1}{a} \sin(p_\mu a)\,,\qquad\qquad\hat p_\mu\equiv \frac{2}{a} \sin\frac{p_\mu a}{2}\,,
\ee
 $A$ is in our case defined in \eqref{A}, and  ($|r|=1$)
          \be\label{Wilsonshiftgen}
W_\pm = \frac{r}{2\,z^2}\,\big({\hat p}_0^2\pm i\,{\hat p}_1^2\big)\,\rho^M z_M\,.
\ee   
We recall that the $U(1)$ symmetry forbids in the original action the presence of bilinears made up of fermions with identical $U(1)$ charge (upper diagonal block entries in \eqref{OF}), and only allows them if some compensating, oppositely charged, field multiplies them (lower diagonal block entries in \eqref{OF}).  The Wilson term $W_\pm$ in \eqref{Wilsonshiftgen} is $U(1)$-neutral, and the breaking of the $U(1)$ symmetry is due to its presence  in the diagonal of \eqref{OFgen}. \\
The values of the discretised (scalar) fields are assigned to each lattice site,  with periodic boundary conditions for all the fields except for antiperiodic temporal boundary conditions in the case of fermions. 

\section{Spectrum of the fermionic operator}
\label{sec:spectrum}

In simpler cases of models with four-fermion interactions~\cite{Catterall:2015zua,Catterall:2016dzf} a choice of  Yukawa terms  similar in spirit to the one described in the previous section turns out to ensure a positive-definite Pfaffian. 
There the relevant operator is real and antisymmetric -- so that its purely imaginary eigenvalues come in pairs $(i\,a,-ia)$ -- \emph{and}  the symmetries of the model ensure that all eigenvalues are also doubly degenerate. One may then define the Pfaffian as the product of eigenvalues with positive imaginary part on the initial configuration. As the simulation progresses, sign flips in the Pfaffian correspond to an odd number of eigenvalues crossing through the origin, but as all eigenvalues are doubly degenerate such sign changes cannot occur. For a system with a  positive-definite Pfaffian the arrow in \eqref{fermionsintegration} is an equivalence, and no sign problem appears.

\begin{figure}[h]
   \centering
 \includegraphics[scale=0.7]{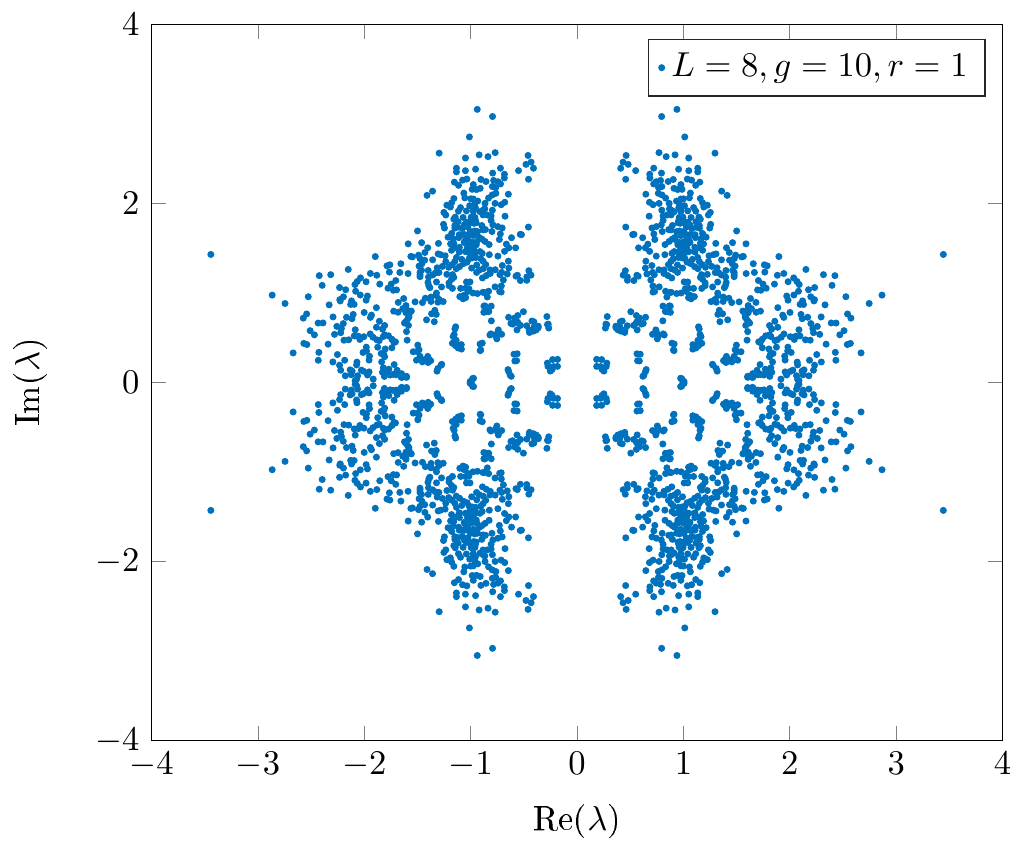}
 \includegraphics[scale=0.7]{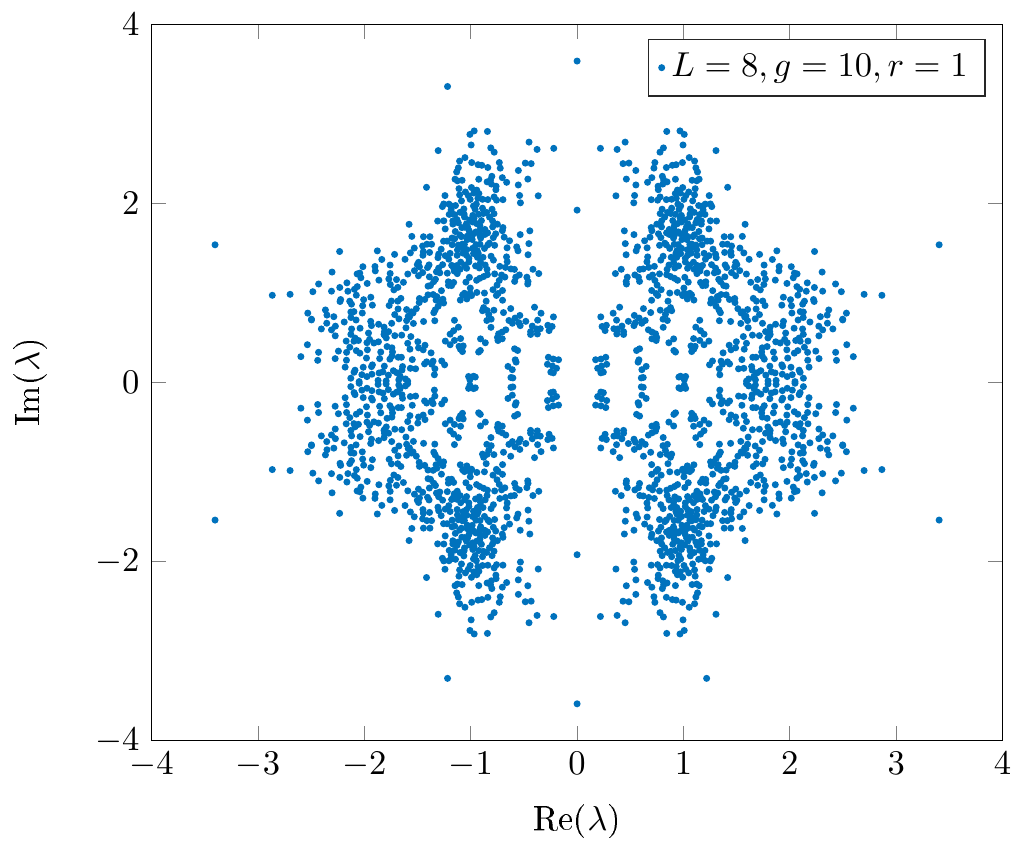}
 \includegraphics[scale=0.7]{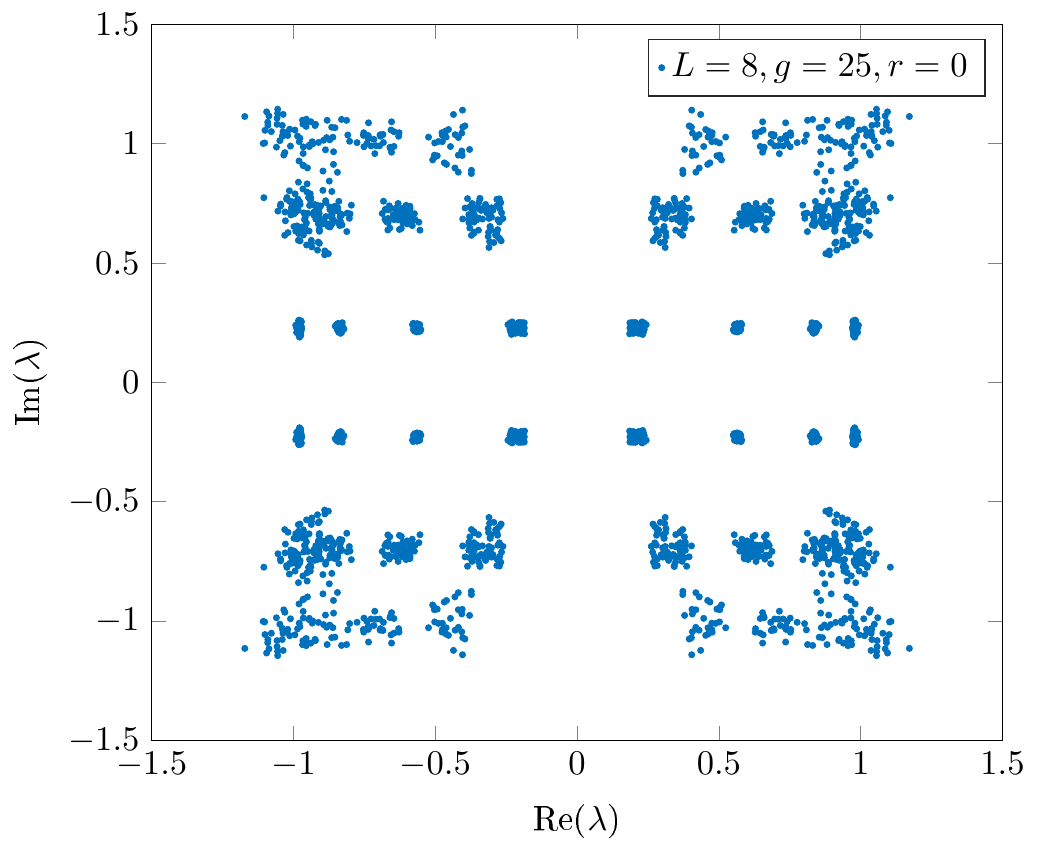}
 \includegraphics[scale=0.7]{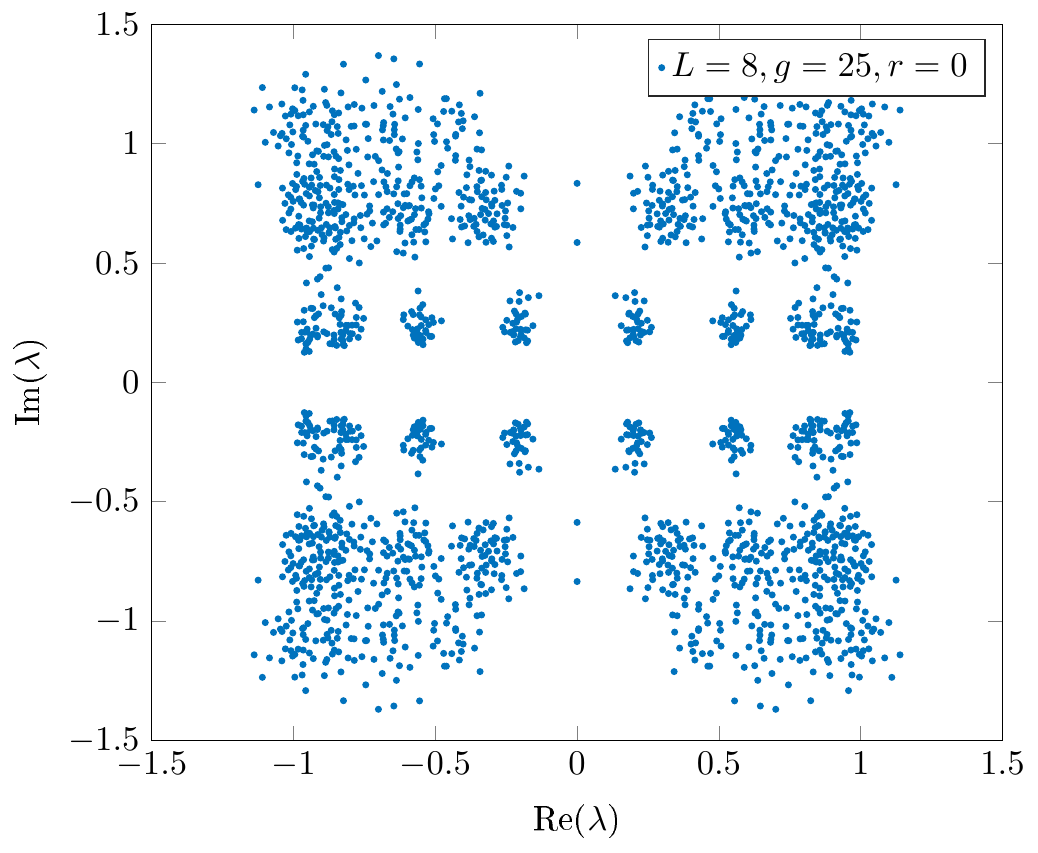}
 \caption{Spectrum of $\hat O_F$, in absence (left diagrams) and presence (right diagrams) of  $A$ (Yukawa-like) terms.}
\label{fig:spectrum_ferm}
\end{figure}

In our case, the fermionic operator $\hat O_F$ is antisymmetric, and satisfies the  constraint (reminiscent of the $\gamma_5$-hermiticity in lattice QCD)~\cite{POS2015,Bianchi:2016cyv}
 \begin{equation}\label{gamma5prop}
\hat O_F^\dagger=\Gamma_5\,\hat O_F\, \Gamma_5\,,
\end{equation}
where $\Gamma_5$ is the following unitary, antihermitian matrix
\be\label{Gamma5}
\Gamma_5=\left(\begin{array}{cccc}
			0 			&  \mathbb{1}			&0& 0 \\
			-\mathbb{1}	&0					&0				&0\\
			0&    0					&0				& \mathbb{1}\\
			0			&0	&-\mathbb{1}		&0
          \end{array}\right)\,,
\qquad   \Gamma_5^\dagger \Gamma_5=\mathbb{1}  \qquad \Gamma_5^\dagger=-\Gamma_5\,.
\ee
The antisymmetry and the property \eqref{gamma5prop} ensure  $\det \hat O_F$ to be \emph{real} and \emph{non-negative}. %In particular, it can be seen that if $\lambda$ is an eigenvalue, then also  $-\lambda^*$ is an eigenvalue. Since OF is skew-symmetric also ?? and ?? 
While the absence of a complex phase allows us to eliminate a systematic error of our previous analysis,  it is not enough to make the Pfaffian positive-definite, implying that the model may still suffer a sign problem.
One can check that -- in the case of generally complex eigenvalues $\lambda$ -- the antisymmetry and the $\Gamma_5$-hermiticity \eqref{gamma5prop} ensure a spectrum characterized by \emph{quartets} $(\lambda, -\lambda^*, -\lambda, \lambda^*)$. %, \colb{see for instance section 3.1 of \cite{peetz2003spectrum} [we should add another ref. perhaps an article of that Thesis] }.
%, which would give a positive-definite Pfaffian. 
One can then define the Pfaffian on the starting configuration as the product $(\lambda\,\lambda^*)$ for each quartet, which would provide sign flips in ${\rm Pf} \hat O_F$. 
However, for purely imaginary or purely real eigenvalues, the disposition in quartets is no longer enforced by \eqref{gamma5prop}  and indeed may not happen, leaving a spectrum of pairs $(\lambda,-\lambda)$ with no degeneracy. A numerical study of the spectrum of $\hat O_F$ appears to indicate that the disposition in quartets would occur if the $A$-terms in~\eqref{OF} -- defining Yukawa-like terms -- were vanishing, see Figure \ref{fig:spectrum_ferm} left, while for $A\neq 0$ (on the right) purely imaginary eigenvalues may appear, with no degeneracy. One should notice that such purely imaginary eigenvalues appear also when auxiliary fields are set to zero  - and thus the only non-vanishing $A$-term is  the one present in the original Lagrangian, before linearization -- suggesting that the sign ambiguity cannot be tamed by a suitably-enough choice of auxiliary fields.  
\begin{figure}[h]
   \centering
 \includegraphics[scale=0.65]{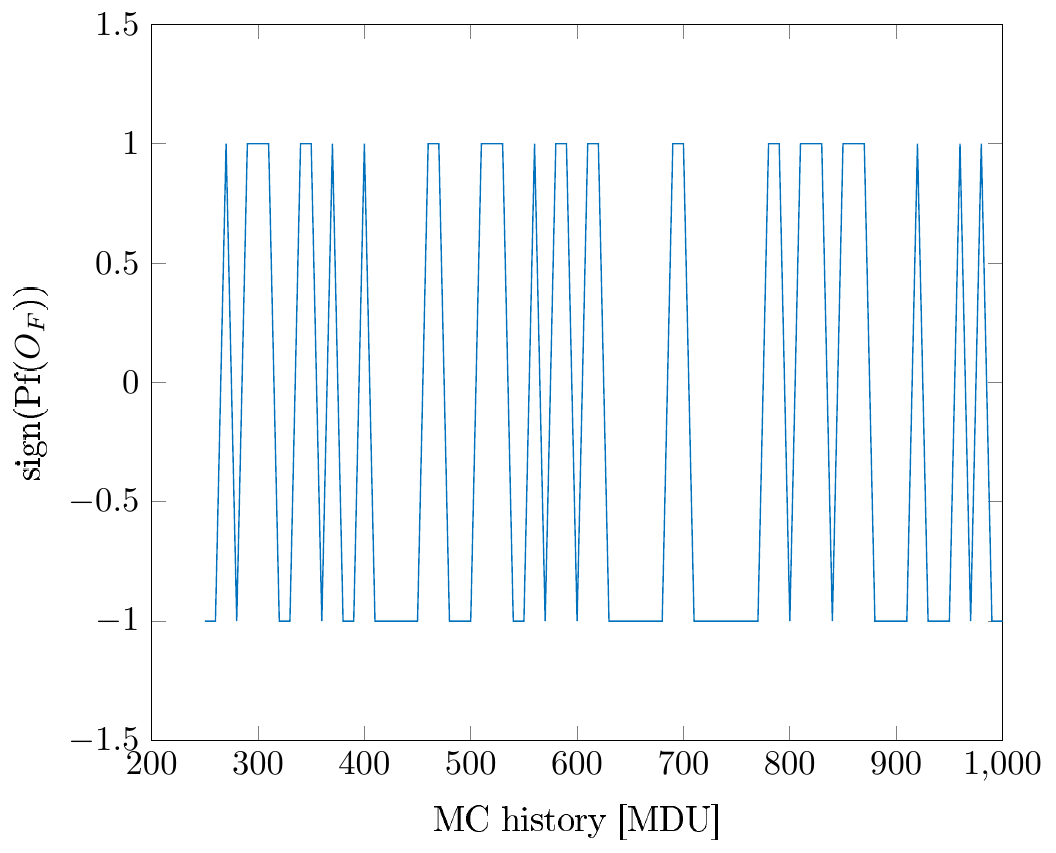}
  \includegraphics[scale=0.65]{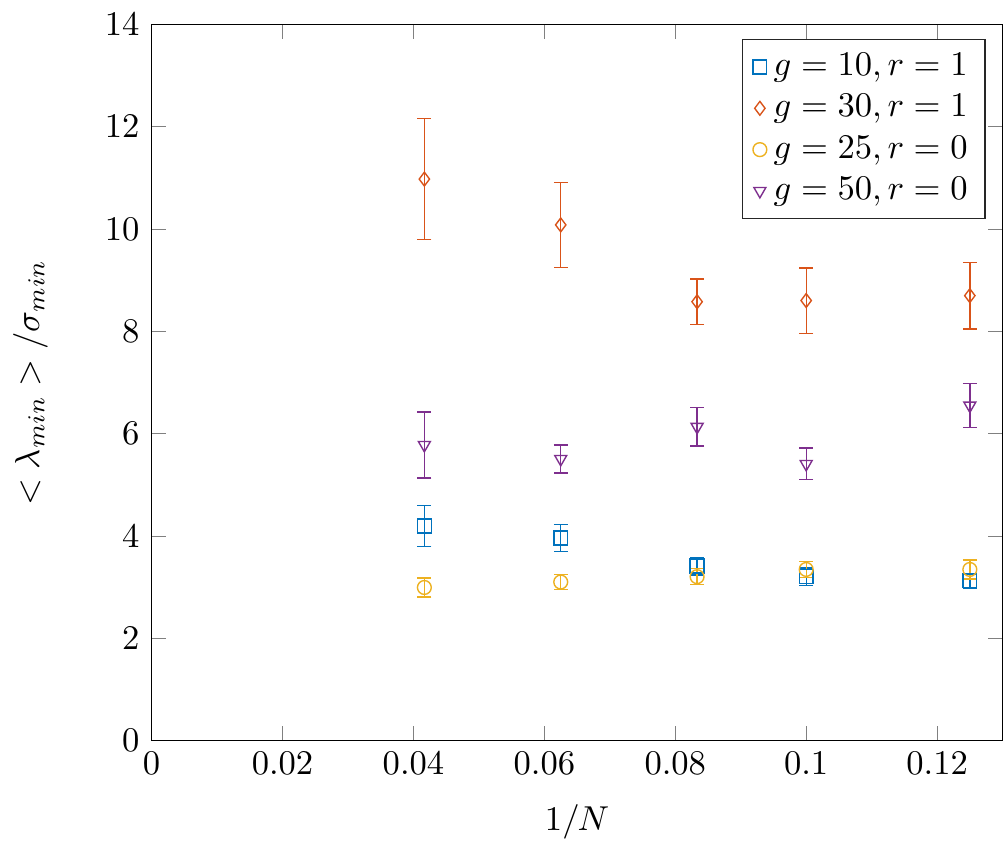}
 \caption{
 %The reweighting factor $W$ accounts for the introduction of a twisted mass term $\mu$ in $O_F$~\cite{Luscher:2008tw} %(via $O_F'=O_F+i\,\mu\,\Gamma_5$) 
 %so to shift the eigenvalues of $O_F$ away from zero ensuring a better convergence of the conjugate gradient solver~\cite{toappear}. 
{\bf Left panel:} Monte Carlo history for  the sign of the Pfaffian of $O_F$ in \eqref{OF} at a value $g=2$ of the coupling. The strong oscillatory behavior indicates a severe sign problem. \\
{\bf Right panel:} The lowest  eigenvalue $\lambda_{\rm min}$ for the squared fermionic operator $O_F^\dagger O_F$  
 appears to be well separated from zero, a statement which then also holds for $O_F$. The variance is defined by $\sigma_{\rm min}^2=\langle \lambda_{\rm min}^2\rangle-\langle \lambda_{\rm min}\rangle^2$. In the region of parameters explored, no zero eigenvalues for $\det O_F$ appear, indicating that for the \emph{real} Pfaffian $\text{Pf} O_F$ no sign flips should occur.}
\label{fig:sign_lambdamin}
\end{figure}

\noindent A sign problem appears already at $g=5$~\cite{Bianchi:2016cyv}, and Fig. \ref{fig:sign_lambdamin} (left panel) shows that the problem becomes severe for values of the coupling $g\sim 2$. 
%
%
%Some interes indication on which region of the parameter space may be free from a sign problem %- and whether in such region information on the non-perturbative behavior of the system is obtainable -- 
It is interesting to look  at the lowest eigenvalue for the squared fermionic operator $\hat  O_F^\dagger \hat O_F$ in a large region of the parameter space. If zero eigenvalues of $\hat  O_F^\dagger \hat O_F$ do not occur for certain values of the parameters, no  zero eigenvalues will occur for $\hat O_F$ as well,  and thus no sign flips for its  Pfaffian. The right panel of Figure~\ref{fig:sign_lambdamin} shows that the smallest eigenvalues of $\hat  O_F^\dagger \hat O_F$ are clearly separated from zero for values of $g\gtrsim 10$. Although not
a proof of their absence, this ``gap'' suggest that sign flips are extremely unlikely. It would be interesting to understand the reason for  this ``gap''. It is also interesting to notice that this region of the parameter space safely includes $g=10$, at which simulations~\cite{Bianchi:2016cyv} appear to detect a non-perturbative behavior~\footnote{We refer here to the measurement of the derivative of the cusp anomaly studied in~\cite{Bianchi:2016cyv}, which show a clear downward behavior - non-perturbative - for $g=10$ and beyond.}. 
%learly, the safest way to determine the region of the parameter space which is free from a sign problem remains a direct evaluation of the 
%\colb{sign flips in the Pfaffian (a continuous function of the eigenvalues) cannot occur at all. This is what is done in Fig. ??, where such distribution appears to be well separated from zero for the region  of couplings explored, $g\gtrsim10$. In fact, it is reasonable to claim that this value  is  testing a  non-perturbative regime of the string sigma-model as $g=10$ measurements  do not belong to the perturbative realm of both observables. The latter is identifiable -- looking at the leading behavior and at the trend of perturbative corrections --  with $f?(g)/4\geq1$ and  $??$, while Fig. ?? shows .. and ??.}
% \begin{figure}[h]
%   \centering
% \includegraphics[scale=0.7]{nsigma-vs-N-1.pdf}
% \caption{The lowest part of the eigenvalue distribution of the quadratic (squared) fermion operator $O_F^\dagger O_F$ is well separated from zero. In the region of parameters explored, no zero eigenvalues for $\det O_F$ occur, which 
% implies no sign flips, and therefore absence of sign problem, for its \emph{real} Pfaffian. }
%\label{fig:lambda_min}
%\end{figure} 

\section{Simulations at finite coupling}
\label{sec:simulations_strong}

We  will now explore the region of the coupling $g< 10$, where a sign problem appears. In addition to the latter, simulations at $g\lesssim5$ run into numerical instabilities due to the non-convergence of the inverter for the fermionic matrix. These instabilities can be traced back to the presence of zero eigenvalues of the fermionic operator,  
%These small eigenvalues correspond to configurations that have a very small weight in the path integral but pose an obstacles to the simulations. 
and may be cured by regularizing the fermionic Pfaffian in a way reminiscent of the twisted-mass reweighting procedure of~\cite{Luscher:2008tw} (see also~\cite{Finkenrath:2013soa}).  Namely, a massive term is added to the fermionic matrix
 to obtain
\be\label{twistedmass}
\tilde O_F=\hat O_F+i\,\mu\,\Gamma_5\,,\qquad\qquad  \tilde O_F {\tilde O_F}^\dagger= \hat O_F {\hat O_F}^\dagger+\mu^2\,\mathbb{1}\,,
\ee
so that $\mu^2\,\mathbb{1}$  shifts the eigenvalues of $ \hat O_F {\hat O_F}^\dagger$ apart from zero.  To compensate for this,  one uses reweighting (see below) and refers to $\mu$ as the reweighting mass parameter. 

Therefore, in this region of the parameter space simulations are not done with the exact string worldsheet action as given by the discretized version of~\eqref{Scuspquadratic} and  \eqref{OF} (in configuration space), but differ due to both the replacement \eqref{fermionsintegration} of the Pfaffian by its absolute value \emph{and} the addition of the  ``twisted mass'' in \eqref{twistedmass}. 
The expectation values $\langle \mathcal{O}\rangle$ of observables in the underlying, target  theory are then obtained from the  expectation values $\langle O \rangle_\text{m}$ in the theory with the modified, positive-definite  fermionic determinant $( \det\big(\tilde O_F {\tilde O_F}^\dagger)+\mu^2\big)^{\frac{1}{4}}$ as follows
\be\label{reweight}
\langle \mathcal{O}\rangle=\frac{\langle\mathcal{O}\,W\rangle_\text{m}}{\langle W \rangle_\text{m}}\,,
\ee
where the total reweighting factor  $W$ reads  in our case~\footnote{Given the exploratory nature of our study, we do not address here a further (so-called RHMC) reweighting factor accounting for the accuracy of the rational approximation for the inversion 
$(\hat O_F \hat O_F^\dagger)^{-\frac{1}{4}}$ in \eqref{fermionsintegration}.}
\be\label{reweight_factors}
W=W_\text{s}\,W_\mu\,,\qquad\qquad W_s=\text{sign}\,\text{Pf}\,\hat O_F\,\qquad\qquad W_\mu=\frac{(\det \hat O_F^\dagger \,\hat O_F)^{\frac{1}{4}}}{\big(\det (\hat O_F^\dagger \,\hat O_F+\mu^2)\big)^\frac{1}{4}}~.
\ee
Below we will investigate two kinds of observables (bosonic and fermionic correlators) and evaluate the reweighting factors exactly, which is feasible in the case of small lattices. We will choose for $\mu$ two different values, and comment on the impact of reweighting on the observables. 

For a part of this paper (see Section~\ref{sec:spectrum} and Section~\ref{sec:reweighting}) we work at finite, relatively small values of $N$, which allows to use exact algorithms for evaluating with reasonable effort  fermion determinants or Pfaffians. In particular, we employ the algorithm in \cite{wimmer2012algorithm} to evaluate the Pfaffian of a matrix without reference to its determinant. All the analysis in Section \ref{sec:observables} the Pfaffian is evaluated stochastically within a rational hybrid Monte Carlo algorithm.
In order to simulate at a point where finite volume effects are small  we fix parameters and  thus the line of constant physics in the bare parameter space
as in \cite{Bianchi:2016cyv}. Namely, in the  space of parameters $(g, N, M)$ -- the dimensionless coupling $g=\frac{\sqrt{\lambda}}{4\pi}$, the number of lattice points $N$ and the dimensionless ``mass'' parameter $M=m\, a$ -- we keep $L \,m\equiv N M =\text{const}\equiv4$. The continuum limit is then taken in this paper via a simple extrapolation to $N\to \infty$. One of the main conclusions of this paper is that this line of constant physics needs to be modified, in view of an infinite renormalization occurring for the fermionic masses. Error bars in the plots below represent statistical errors and include
effects of auto-correlation in the Monte Carlo data \cite{Wolff:2003sm}.

% 
%
%
 %
%  \begin{figure}[t]
%   \centering
%   \includegraphics[scale=.7]{so6/plots/history-Cx-L16-g10-Lm4.pdf}
%    \includegraphics[scale=.7]{so6/plots/history-action-L16-g10-Lm4.pdf}
 %\caption{Monte Carlo histories for the  correlator $\langle x\,x^*\rangle$ at time separation $T/4$ and for $\langle S_{\rm cusp}\rangle$, at $g=10$ and $L/a=16$, in terms of Molecular Dynamic Units (MDU). 
%The HMC produces a series of bosonic field configurations, on each of them the observable is evaluated and plotted here for the same series  at the given parameters. The fact that  successive configurations produced by the RHMC are statically correlated might lead to strong so-called auto-correlations in the data, which would appear in these plots as fluctuations with long periods. As one can see, the histories presented here do not suffer from such long fluctuations, and sample well the observables under investigation.}
%\label{fig:MChistories}
%\end{figure} 
%
%
%
%
%
%
%
%
%
%
\begin{table}[H]
\centering
\begingroup
\renewcommand*{\arraystretch}{0.9}
\begin{tabular}{cccccccc}
\toprule
$g$ & $T/a\times L/a$ & $Lm$ & $am$ & $\mu$ \\
\midrule
  2 & $16 \times   8$ &  4  & 0.50000 & $0.01$ \\
\midrule
  5 & $16 \times   8$ &  4  & 0.50000 & $0.01$ \\
  5 & $16 \times   8$ &  4  & 0.50000 & $0.02$ \\
\midrule
 10,20,25,30,50,100 & $16 \times   8$ &  4  & 0.50000 & $0.0$ \\
  & $20 \times   10$ &  4  & 0.40000 & $0.0$ \\
  & $24 \times   12$ &  4  & 0.33333 & $0.0$ \\
  & $32 \times   16$ &  4  & 0.25000 & $0.0$ \\
  & $48 \times   24$ &  4  & 0.16667 & $0.0$ \\
  & $64 \times   32$ &  4  & 0.12500 & $0.0$ \\
\bottomrule
\end{tabular} 
\endgroup
\caption{
The parameters of our simulations are the coupling $g$, the temporal ($T$) and spatial ($L$) extent of the lattice in units of the lattice spacing $a$. The mass parameter $am$ is given by
the fixing the combination $Lm=4$. The reweighting parameter $\mu$ is non-zero only for $g<10$.}
\label{t:runs}
\end{table}
Table \ref{t:runs}  collects the parameters of the simulations here presented. Configurations are generated by the standard Rational Hybrid Monte Carlo (RHMC) algorithm~\cite{RHMC1,RHMC2}, with a  rational approximation of degree $15$ for the inverse fractional power in \eqref{fermionsintegration}.

%In Fig.\ref{fig:MChistories} we show examples of Monte Carlo histories for our two main observables - the correlator $\langle x^* x\rangle$  and the action $\langle S_{\rm cusp}\rangle$. 
%We determined  auto-correlation times of the observables and included their effect in the error analysis~\cite{Wolff:2003sm}. 
%Multiple points at the same value of $g$ and $N$ in Fig. \ref{fig:correlator} (left panel), Fig. \ref{fig:action_div} and Fig. \ref{fig:action_fin_N2}
%-- and similarly in Fig. \ref{fig:correlator_break} (left panel), Fig. \ref{fig:action_div_break} and Fig. \ref{fig:action_fin_N2_break} -- indicate multiple replica.  

%In this setup we present below our main observables here, the correlator of the bosonic $x$ field and the correlators of the fermionic fields.

\subsection{Observables}
\label{sec:observables}

\subsubsection{The $\langle xx^{*}\rangle$ correlator}
\label{sec:correlator}

We use the new linearization of the (discretized) Lagrangian \eqref{Scuspquadratic} with \eqref{A}-\eqref{Wilsonshiftgen} to repeat the analysis for the mass of the bosonic field $x$ in section 4.1 of \cite{Bianchi:2016cyv}. Here, we defined the timeslice correlation function on the lattice at given time interval $t$
\be\label{C_x}
C_{x} (t;k)\equiv \sum_{s,s^{'}} e^{-i k (s_1-s_2)} G_{x} (t,s,0,s')
\ee
from the \emph{connected} two-point function
\be\label{G_x}
G_{x} (t,s,t',s')\equiv \braket{x(t,s) x^*(t',s')}_{\rm c} = \braket{x(t,s) x^*(t',s')} - \braket{x(t,s)} \braket{x^*(t',s')}\,.
\ee
The subtraction of the one-point functions is irrelevant in the continuum, where the $U(1)$ invariance implies $\braket{x}=\braket{x^*}=0$, but is crucial on the lattice, where the Wilson term breaks this symmetry. 
%~\footnote{ The subtraction in  \eqref{G_x} is implicit in (4.4) of~\cite{Bianchi:2016cyv}.}. 
The non-trivial, and linearly divergent, one-point functions of $\tilde{x},\,\tilde{x}^*$  are calculated at leading order in lattice perturbation theory in Appendix~\ref{app:onepoint}. In Fig.~\ref{fig:x-vev} we show the plot of $\langle x\rangle$ for several values of $g$ and $N$. 

\begin{figure}[H]
   \centering
           \includegraphics[scale=0.6]{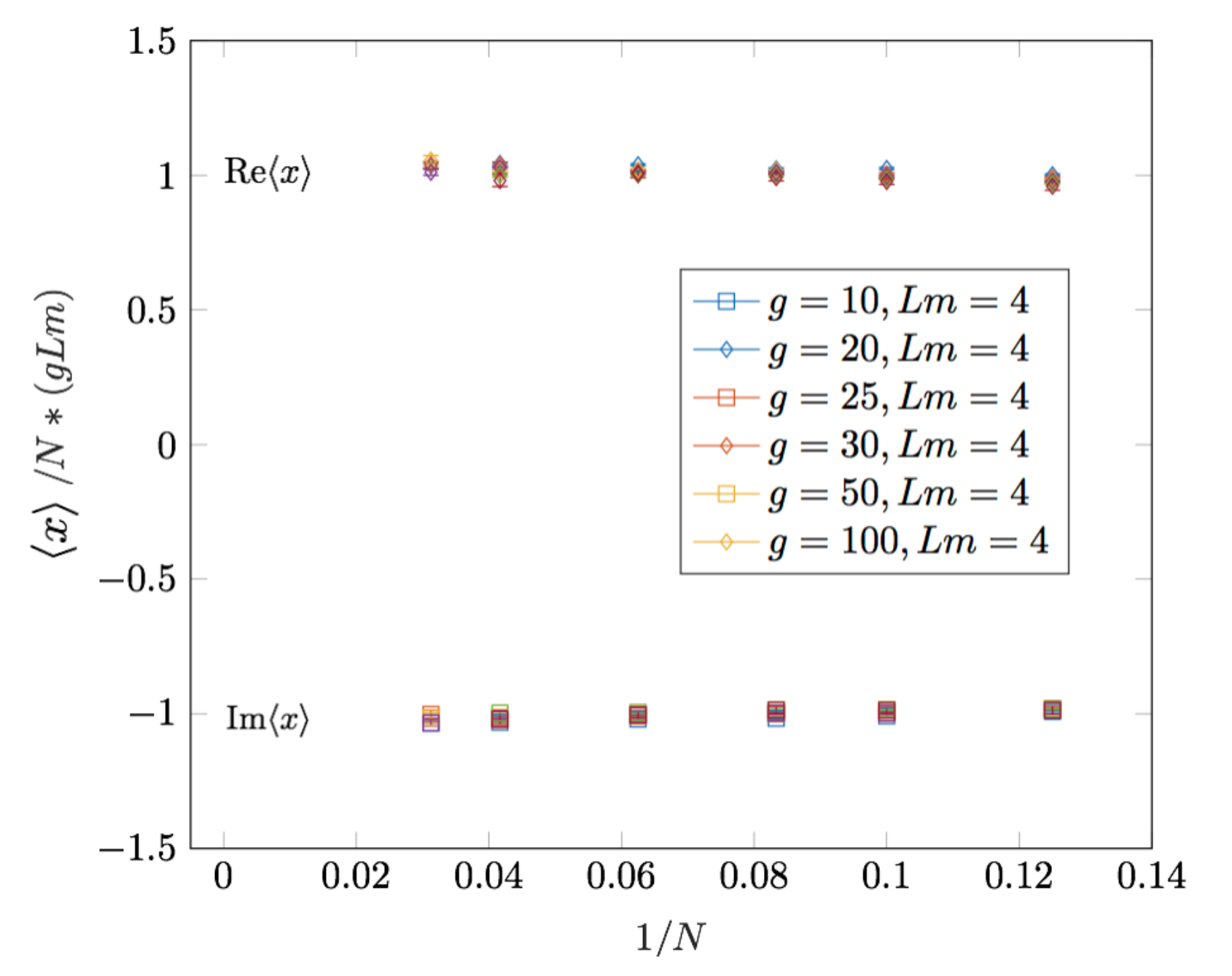}        
 \caption{Plot of the real and imaginary part of $\langle x\rangle$ for several values of $g$ and $N$. The vacuum expectation value is normalized by $N/(gLm)$, namely the perturbation theory result \eqref{vevx} at $\mathcal{O}(1/g)$, and therefore the constant behavior visible in the flatness shows for $\langle x\rangle$ a divergence which is \emph{linear} in $N$.}
             \label{fig:x-vev}
\end{figure}

The exponential fall-off of the timeslice correlator for large interval $t$ and zero momentum defines the physical mass of the fluctuation $x$
\begin{gather}\label{correlator_groundstate}
C_x(t;\,0)~ \stackrel{t\gg1}{\sim} ~e^{- t\, {m_x}_{\rm LAT}}\,.
\end{gather}
%While this is already noted in (4.7) \cite{Bianchi:2016cyv}, such asymptotic behaviour is a good approximation as along as the time separation $t$ is much smaller than the lattice size $T$. 
On the lattice the periodic boundary condition on the field $x$ in the time direction imposes the relation $C_{x}(t) = C_{x}(T-t)$, which means that \eqref{correlator_groundstate} is rather
\begin{equation}
C_x(t;\,0)~ \stackrel{t\gg1}{\sim} e^{-tm_{x \rm LAT}}+e^{-(T-t)m_{x \rm LAT}} \,.
%= \cosh\left(\left(\frac{T}{2}-t\right) m_{x\rm LAT}\right).
\end{equation}
The value of the physical mass is measured, on the lattice,  from the limit of an effective mass $m^\textrm{eff}_{x}$ for fixed lattice time extension $T$
\begin{gather}
\label{measured_m_eff}
{m_x}_{\rm LAT}=\lim_{T,\,t\to\infty}m^\textrm{eff}_{x}\,.
\end{gather}
We estimate the latter by fitting the timeslice correlator $C_{x}(t;0)$ with a double exponential 
\begin{equation}\label{fit-mx}
A \Big[e^{-t m^\textrm{eff}_{x}}+e^{-(T-t)m^\textrm{eff}_{x}}\Big]
%A\cdot e^{-Tm_{x}^{\rm eff}/2}\cdot\text{cosh\,}\left(\left(\frac{T}{2}-t\right) m_{x}^{\rm eff}\right).
\end{equation}
on the interval $1 \ll t \ll T$. The overall factor $A$ is irrelevant; measurements of $m^\textrm{eff}_{x}$ improve when $T=2L$ and data points at $t\sim{T}/{2}$, which are affected by the largest relative errors, are discarded. A major source of uncertainty comes from the estimate of the one-point functions in \eqref{G_x}, which is reduced as follows. Denoting the Fourier component of $x$ at zero spatial momentum by 
\begin{equation}
\tilde{x}(t)\equiv \sum\limits_{s} x(t,s)
\end{equation}
and splitting the field $x$ into real $x_{\rm R}$ and imaginary part $x_{\rm I}$, the connected timeslice correlator \eqref{C_x} takes the form
\begin{flalign}\label{rel1}
\langle \tilde{x}(t) \tilde{x}^{*}(0) \rangle_{\rm c} &= \langle\tilde{x}_{\rm R}(t)\tilde{x}_{\rm R}(0)\rangle + \langle \tilde{x}_{\rm I}(t)\tilde{x}_{\rm I}(0)\rangle
 - \langle \tilde{x}_{\rm R}(t)\rangle\langle \tilde{x}_{\rm R}(0)\rangle - \langle \tilde{x}_{\rm I}(t)\rangle\langle \tilde{x}_{\rm I}(0)\rangle \\
 &  +i \left( \langle\tilde{x}_{\rm I}(t)\tilde{x}_{\rm R}(0)\rangle - \langle\tilde{x}_{\rm R}(t)\tilde{x}_{\rm I}(0)\rangle \right) \,.
 \nonumber
\end{flalign}
The second line vanishes due to translational and time-reversal invariance. In Appendix \ref{app:onepoint} we show that it holds
\be\label{vevrvevim}
\langle \tilde{x}_{\rm R}\rangle = - \langle \tilde{x}_{\rm I}\rangle\,,
\ee
while the relations~\footnote{The second equation follows from the first for translational invariance.}
\begin{flalign}\label{disctoconn}
\langle \tilde{x}_{\rm R}(t)\, \tilde{x}_{\rm I}(0)\rangle = \langle \tilde{x}_{\rm R}(t)\rangle \langle\tilde{x}_{\rm I}(0)\rangle\,,
~~~~
\langle \tilde{x}_{\rm I}(t)\, \tilde{x}_{\rm R}(0)\rangle = \langle \tilde{x}_{\rm I}(t)\rangle \langle\tilde{x}_{\rm R}(0)\rangle\,
\end{flalign}
are observed to hold within numerical precision. These last two equations allow us to trade the disconnected pieces in \eqref{rel1} with connected ones, e.g.  $\langle \tilde{x}_{\rm R}(t)\rangle\langle \tilde{x}_{\rm R}(0)\rangle=-\langle \tilde{x}_{\rm R}(t)\tilde{x}_{\rm I}(0)\rangle$, which  brings \eqref{rel1} into the form
\begin{equation}\label{eq: conn_corr}
C_{x}(t;\,0) = \langle\tilde{x}_{\rm R}(t)\tilde{x}_{\rm R}(0) + \tilde{x}_{\rm I}(t)\tilde{x}_{\rm I}(0) + \tilde{x}_{\rm R}(t)\tilde{x}_{\rm I}(0) + \tilde{x}_{\rm I}(t)\tilde{x}_{\rm R}(0) \rangle\,
\end{equation}
and substantially reduces the statistical error.
\begin{figure}[H]
   \centering
                \includegraphics[scale=1]{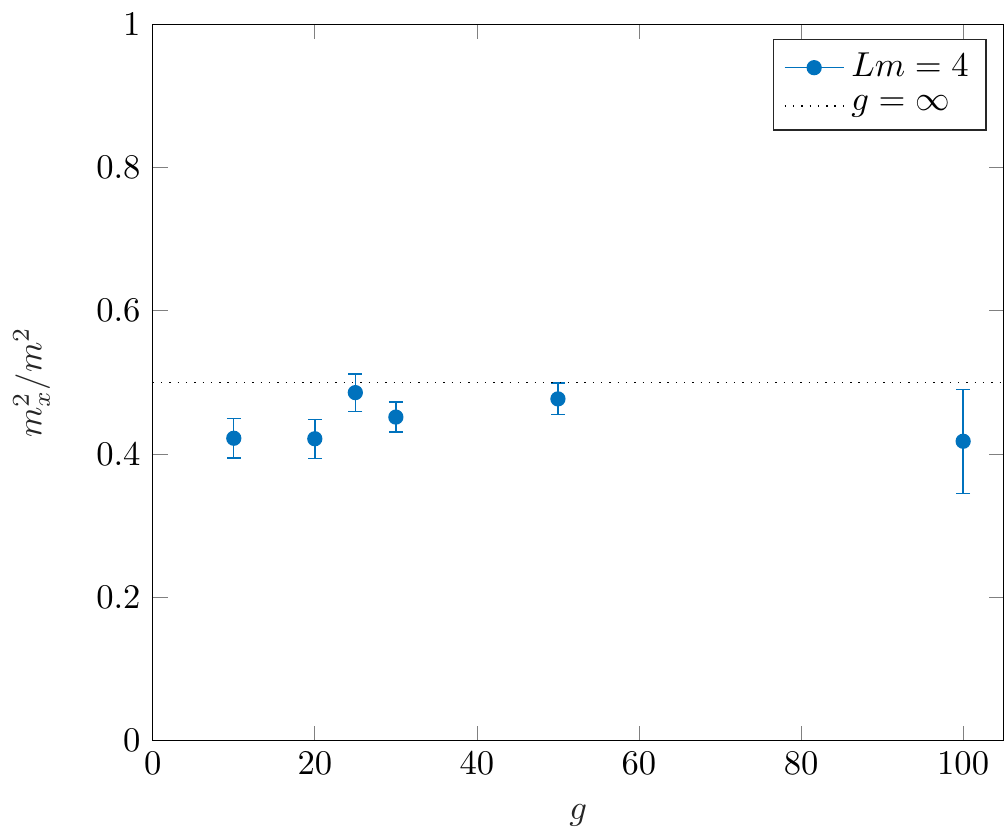}
                    \caption{Continuum values for the measured $x$ mass versus $g$ (blue dots). The extrapolation of
                    the values at finite lattice spacing to the continuum limit is performed as
                    in \cite{Bianchi:2016cyv}. The dotted line is the $g\to\infty$ limit of the continuum prediction.}
             \label{fig:mass-x-vs-g}
\end{figure}
Figure~\eqref{fig:mass-x-vs-g} shows the measured $x$ mass, as extrapolated in the continuum from~\eqref{measured_m_eff}. The  estimate is  consistent with the large $g$, continuum prediction $m^2_{ x}(g)=\frac{m^2}{2}\,\Big(1-\frac{1}{8 \,g}+\mathcal{O}(g^{-2})\Big)~$ (see discussion in~\cite{Bianchi:2016cyv}). As already noticed in~\cite{Bianchi:2016cyv}, there appears to be no infinite renormalization occurring for $m^2_x$. As we will see below in Section \eqref{sec:fermions}, however, this is not the case for the fermionic masses, implying that eventually the bare parameter $m$ will have to be tuned to adjust for it and the continuum limit will have to be reformulated.

\subsubsection{The fermionic correlators}
\label{sec:fermions}

The fermionic generating functional on the lattice is defined by 
\begin{flalign}\label{Z_F}
 Z^{\textrm{LAT}}_F[J] &
 \equiv \int [D\psi]\,  e^{\frac12 \sum_{t, s, t',s'}\psi^T (t,s) \,O_F (t,s, t',s') \,\psi(t',s')+\sum_{t,s}\psi^T(t, s) \,J(t, s)}\\
& =\Pf(O_F)\, e^{\frac12 \sum_{t, s, t', s'} J^T(t, s)\, O_F^{-1}(t, s,t', s')\, J(t',s')}\nonumber
\end{flalign}
and evaluated for a given configuration of the bosonic fields.  $J$ is a 16-component vector of Grassmann-valued source fields conjugated to the fermionic field $\psi=({\theta}^i, { \theta}_i, {\eta}^i, {\eta}_i)$ with $i,j=1,...4$, and sums run over the lattice sites indexed by $t=1,... 2N$ and $s=1,... N$. Fermionic two-point functions are obtained differentiating \eqref{Z_F} with respect to $J^{\hat{i}}$ with $\hat{i},\hat{j}=1,...16$
\begin{equation}
\left. \frac{\partial}{\partial {J^{\hat{i}} (t,s)}} \frac{\partial}{\partial {J^{\hat{j}} (t',s')}}
 Z^{\textrm{LAT}}_F[J]
\right|_{J=0}= \Pf(O_F)\,[O_F^{-1}(t,s,t',s')]_{\hat{i}\hat{j}}
\end{equation}
and integrated over the bosonic fields to obtain the relation
%~\footnote{It delivers the connected part of the correlator because one-point functions are zero by definition of Grassmann integral.}
\begin{equation}
G_{\psi_{\hat{i}} \psi_{\hat{j}}} (t,s,t',s') \equiv \braket{\psi_{\hat{i}}(t,s) \psi_{\hat{j}}(t',s')}=\braket{[O_F^{-1}(t,s,t',s')]_{\hat{i}\hat{j}}}\,.
\end{equation}
For the various components we extract the following two-point functions
\begin{align}\label{list_ferm_2pt}
 G_{\theta^i \theta^j} (t,s,t',s')&=\braket{[O_F^{-1}(t,s,t',s')]_{i,j}}\,, ~~~~~~~~
   G_{\theta^i \theta_j} (t,s,t',s')=\braket{[O_F^{-1}(t,s,t',s')]_{i,j+4}}\,,\no\\
  G_{\eta^i \eta^j} (t,s,t',s')&=\braket{[O_F^{-1}(t,s,t',s')]_{i+8,j+8}}\,, ~~
   G_{\eta^i \eta_j} (t,s,t',s')=\braket{[O_F^{-1}(t,s,t',s')]_{i+8,j+12}}\,,\no\\
  G_{\theta^i \eta^j}(t,s,t',s')&=\braket{[O_F^{-1}(t,s,t',s')]_{i,j+8}}\,, ~~~~~
    G_{\theta^i \eta_j} (t,s,t',s')=\braket{[O_F^{-1}(t,s,t',s')]_{i,j+12}}\,.
\end{align}
In analogy with \eqref{G_x}, to evaluate the mass we define timeslice correlators of fermionic fields on the lattice as
\be \label{C_ferm}
C^{\textrm{LAT}}_{\psi^{\hat{i}}\psi^{\hat{j}}}(t;k) = \sum_{s_1, s_2} e^{-i k (s_1-s_2)} G_{\psi^{\hat{i}}\psi^{\hat{j}}} (t,s_1,0,s_2)\,
\ee
and project on the zero spacial momentum $k=0$.
 
As usual, it is instructive to start considering the perturbative region. At large $g$, the inverse of the fermionic operator \eqref{OF} in momentum-space representation reads 
\begin{flalign}\label{Kinverse}
 & K_F^{-1}(p_0,p_1) 
 =\left[\textrm{det}{K}_{F}(p_0,p_1) \right]^{-1/8} {\hat{K}_{F}^{\dagger}(p_0,p_1)}
\end{flalign}
where
\begin{gather}\label{detKF}
\left[\textrm{det}{K}_{F}(p_0,p_1) \right]^{1/8}=\mathring{p_0} ^{2}+\mathring{p_1} ^{2}+\frac{m^{2}}{4}+\frac{a^2 \,r^{2}}{4}\left(\hat{p}_{0}^{4}+\hat{p}_{1}^{4}\right)
\end{gather}
and
\begin{gather}
{\hat{K}_{F}^{\dagger}(p_0,p_1)}
=
\left(\begin{array}{cccc}\label{Kdagger}
\frac{r}{2}\left(\hat{p}_{0}^{2}-i\hat{p}_{1}^{2}\right)\rho_{M}^{\dagger}u^{M} & {-\mathring{p_0} \mathbb{1}} & {-\left(\mathring{p_1} -i\frac{ma}{2}\right)\rho_{M}^{\dagger}u^{M}} & 0\\
{-\mathring{p_0} \mathbb{1}} & -\frac{r}{2}\left(\hat{p}_{0}^{2}+i\hat{p}_{1}^{2}\right)\rho_{M}u^{M} & 0 & {-\left(\mathring{p_1} -i\frac{ma}{2}\right)\rho_{M}u^{M}}\\
{\left(\mathring{p_1} +i\frac{ma}{2}\right)\rho_{M}^{\dagger}u^{M}} & 0 & \frac{r}{2}\left(\hat{p}_{0}^{2}+i\hat{p}_{1}^{2}\right)\rho_{M}^{\dagger}u^{M} & {-\mathring{p_0} \mathbb{1}}\\
0 & {\left(\mathring{p_1} +i\frac{ma}{2}\right)\rho_{M}u^{M}} & {-\mathring{p_0} \mathbb{1}} & -\frac{r}{2}\left(\hat{p}_{0}^{2}-i\hat{p}_{1}^{2}\right)\rho_{M}u^{M}
\end{array}\right)\,
\end{gather}
and  we temporarily reinstated the lattice spacing $a$. The inverse Fourier transform of the matrix entries of \eqref{Kinverse} over the time-like momentum component
\be
C_{\psi_{\hat{i}}\psi_{\hat{j}}}(t,p_1) = \frac{a}{g} \int_{-\infty}^{\infty} dp_0 \, e^{i p_0 t} [K_F^{-1} (p_0, p_1)]_{\hat{i}\hat{j}}
\ee
yields the following analytic predictions for the timeslice correlators \eqref{C_ferm} at $g\gg 1$  
\begingroup \allowdisplaybreaks
\begin{eqnarray}\label{fermions-an-beg}
&&\!\!\!\!\!C_{\theta^{i}\theta^{j}}(t;0)=C_{\eta^{i}\eta^{j}}(t;0)= \frac{-\pi\,u^{M}\left(\rho_{M}^{\dagger}\right)^{ij}}{g\,\sqrt{4-m^{2}a^{2}r^{2}}}\left[{\bar V}_{-}\exp\left(-\frac{t}{ar}{\bar V}_{-}\right)-{\bar V}_{+}\exp\left(-\frac{t}{ar}{\bar V}_{+}\right)\right] \\
&&\!\!\!\!\!C_{\theta_{i}\theta_{j}}(t;0)=C_{\eta_{i}\eta_{j}}(t;0)= \frac{\pi\,u^{M}\left(\rho_{M}\right)_{ij}}{g\,\sqrt{4-m^{2}a^{2}r^{2}}}\left[{\bar V}_{-}\exp\left(-\frac{t}{ar}{\bar V}_{-}\right)-{\bar V}_{+}\exp\left(-\frac{t}{ar}{\bar V}_{+}\right)\right] \\
&&\!\!\!\!\!C_{\theta^{i}\theta_{j}}(t;0)=C_{\theta_{i}\theta^{j}}(t;0)=\\
&&\!\!\!\!\!=C_{\eta^{i}\eta_{j}}(t;0)= C_{\eta_{i}\eta^{j}}(t;0) =\frac{-2\pi i\delta_{ij}}{g\,\sqrt{4-m^{2}a^{2}r^{2}}}\left[\exp\left(-\frac{t}{ar}{\bar V}_{-}\right)-\exp\left(-\frac{t}{ar}{\bar V}_{+}\right)\right]\\
&&\!\!\!\!\!{C_{\theta^{i}\eta^{j}}(t;0)=C_{\eta^{i}\theta^{j}}}(t;0)  =\frac{ima\pi r\,u^{M}\left(\rho_{M}^{\dagger}\right)^{ij}}{g\,\sqrt{4-m^{2}a^{2}r^{2}}}\left[\frac{\exp\left(-\frac{t}{ar}{\bar V}_{-}\right)}{{\bar V}_{-}}-\frac{\exp\left(-\frac{t}{ar}{\bar V}_{+}\right)}{{\bar V}_{+}}\right]\\
&&\!\!\!\!\!{C_{\theta_{i}\eta_{j}}(t;0)=C_{\eta_{i}\theta_{j}}}(t;0) =\frac{ima\pi r\,u^{M}\left(\rho_{M}\right)_{ij}}{g\,\sqrt{4-m^{2}a^{2}r^{2}}}\left[\frac{\exp\left(-\frac{t}{ar}{\bar V}_{-}\right)}{{\bar V}_{-}}-\frac{\exp\left(-\frac{t}{ar}{\bar V}_{+}\right)}{{\bar V}_{+}}\right]\\\label{fermions-an-end}
&&\!\!\!\!\!C_{\theta^{i}\eta_{j}}(t;0)=C_{\theta_{i}\eta^{j}}(t;0)
=C_{\eta^{i}\theta_{j}}(t;0)=C_{\eta_{i}\theta^{j}}(t;0)=0 
\end{eqnarray}
\endgroup
with
\begin{gather}
{\bar V}_\pm= \sqrt{2\pm\sqrt{4-m^{2}a^{2}r^{2}}}\,.
\end{gather}
In the continuum limit ($a\to 0$)  ${\bar V}_+ = 2+\mathcal{O}(a^2)$  and ${\bar V}_- \sim \textstyle\frac{a \,m\,r}{2}$. Therefore,  of the exponentials $\exp\left(-\frac{t}{ar}{\bar V}_{\pm}\right)$, only the ones with ${\bar V}_-$ survive. The propagators in the first two lines above vanish in the limit, while the remaining (non-vanishing) correlators reduce to a single exponential
\begingroup \allowdisplaybreaks
\begin{flalign}\label{diagonal}
C_{\theta^{i}\theta_{j}}(t;0)=C_{\theta_{i}\theta^{j}}(t;0)=C_{\eta^{i}\eta_{j}}(t;0)=C_{\eta_{i}\eta^{j}}(t;0)
&=
-\frac{\pi i}{g}\,\delta^i_{j}\exp\left(-\frac{tm}{2}\right)
\\
C_{\theta^{i}\eta^{j}}(t;0)=C_{\eta^{i}\theta^{j}}(t;0)
&=
\frac{i\pi}{g}\,u^{M}\left(\rho_{M}^{\dagger}\right)^{ij}\exp\left(-\frac{tm}{2}\right)
\\
C_{\theta_{i}\eta_{j}}(t;0)=C_{\eta_{i}\theta_{j}}(t;0)
&=
\frac{i\pi}{g}\,u^{M}\left(\rho_{M}\right)_{ij}\exp\left(-\frac{tm}{2}\right)\,,
\end{flalign}
\endgroup
in agreement with the continuum results~\cite{Giombi}. 
 Notice that the prediction based on the integrability of the model (namely, the study of the dispersion relations for these modes~\cite{Basso:2010in} via the asymptotic Bethe Ansatz) is that that the masses of the fermionic fields should not get renormalized, holding their value $m/2$ for all values of the coupling. 

For our measurements we consider the diagonal correlators \eqref{diagonal}. In fact, to reduce the variance we use the $SU(4)\sim SO(6)$ symmetry and look at their averaged values
\begin{align}
C_{\theta\theta}(t)&=\frac{1}{8}\sum_{i,j} \Big[\,C_{\theta^{i}\theta_{i}}(t) +C_{\theta_i\theta^{i}}(t)  \, \Big]\,,\\
C_{\eta\eta}(t)&=\frac{1}{8}\sum_{i,j} \Big[\,C_{\eta^{i}\eta_{i}}(t)+C_{\eta_i\eta^{i}}(t)\, \Big]\,.
\end{align}
%The analysis above, see equations \eqref{fermions-an-beg}-\eqref{fermions-an-end}, \colb{suggests that for 
and at the sum $C_{\text{sum}}= ({C}_{\theta\theta}+C_{\eta\eta})/2$. The discussion above suggests to fit the Monte Carlo data to a single exponential decay, similar to \eqref{fit-mx}. Such fits were tried but rejected because of their large $\chi^2$ values of the chi-squared test. However, as will become clear below, the data from finite lattices with temporal extent $T$ and anti-periodic boundary conditions can be fitted to the function
\begin{eqnarray}\label{mf-fit}
%\text{For $C_{\theta\theta}$ and $C_{\eta\eta}$ four parameters: $A$, $B$ and $V\pm$} \\
%C(t)=A\, e^{- t\,V_-} +B\,e^{- t\,V_+} + (t\to T -t)\\\nonumber
%\text{For the sum $C_{\text{sum}}= {C}_{\theta\theta}+C_{\eta\eta}$ three parameters: $A$, and $V\pm$} \\
C_{\text{sum}}(t)\sim \, \,e^{- t\,V_-} +\,e^{- t\,V_+} + (t\to T -t)\,.
\end{eqnarray}

\begin{figure}[H]
%   \centering
                 \includegraphics[scale=0.75]{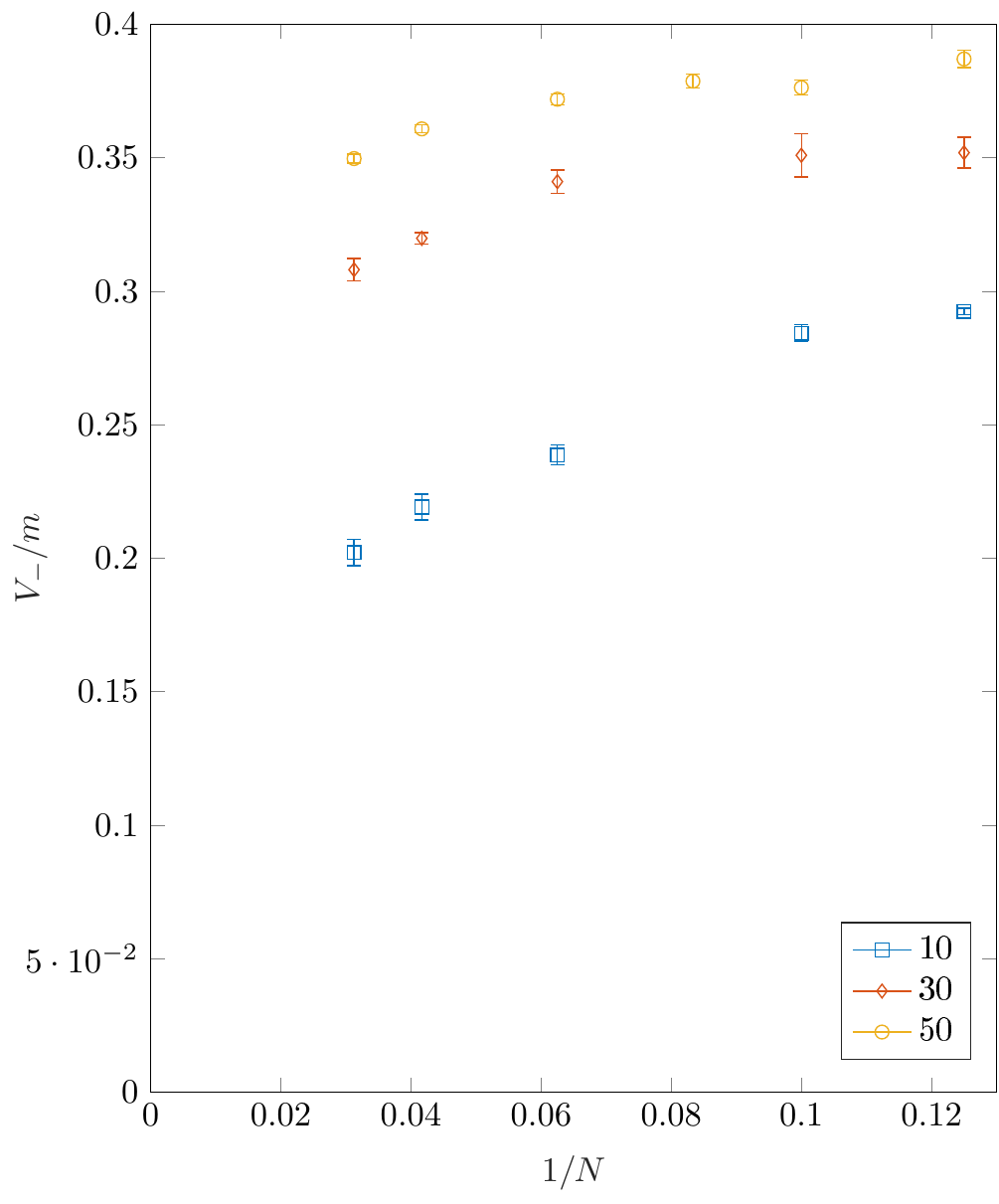}
          % \hspace{0.1cm}
                  \includegraphics[scale=0.75]{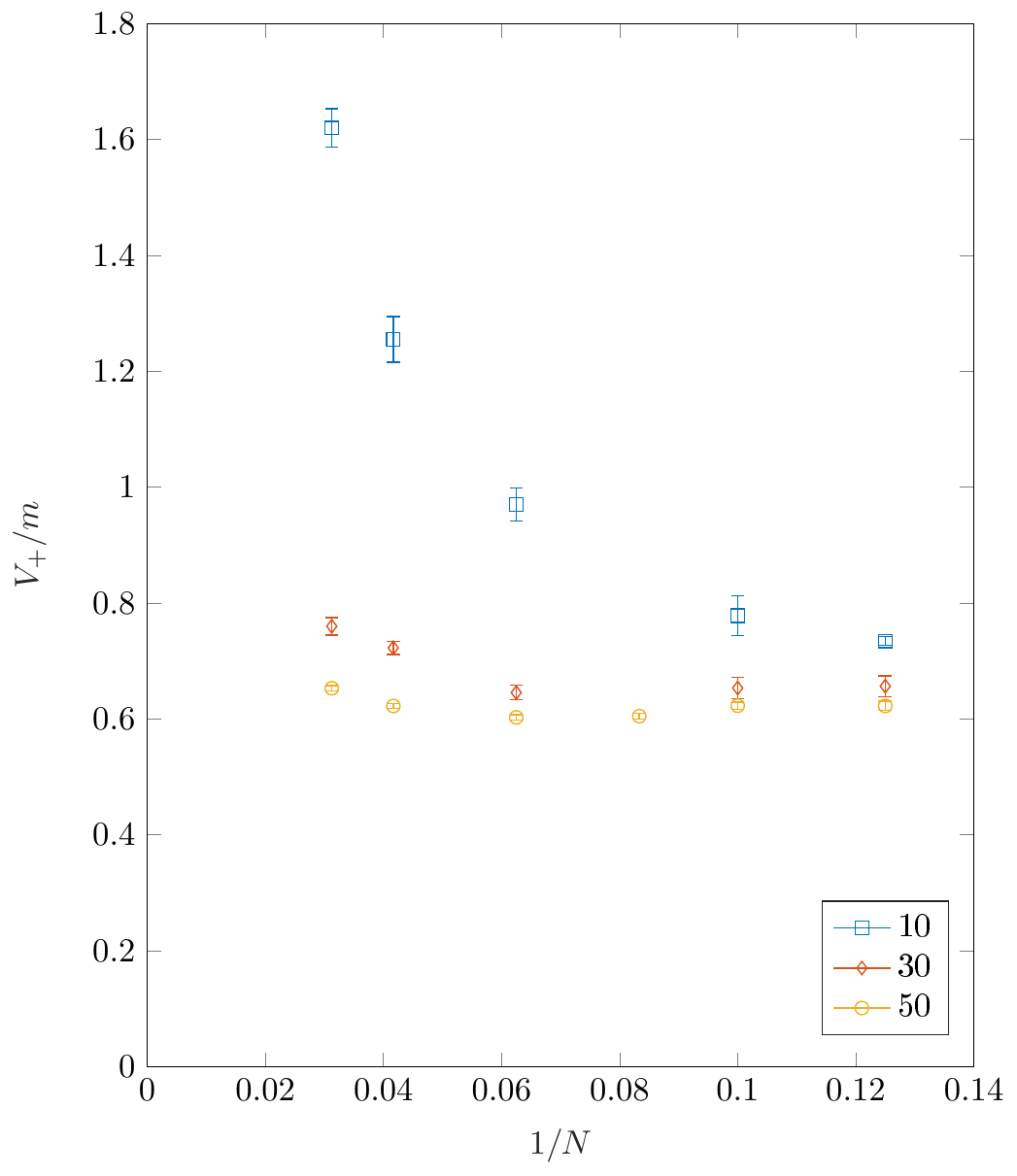}
          % \hspace{0.1cm}
           % \hspace{0.1cm}
          %      \includegraphics[scale=0.75]{VpmMC_VpmPT.pdf}        
                %  \caption{ } 
        % \hspace{0.1cm}         
            \caption{The exponential decays resulting from the fit of the Monte Carlo data for the fermionic correlators $C_{\text{sum}}$ to \eqref{mf-fit} for $g=10,30,50$ and various values of $N$.
            }
             \label{fig:fermioncorrelators}
\end{figure}

\begin{figure}[H]
   \centering
                         \includegraphics[scale=1]{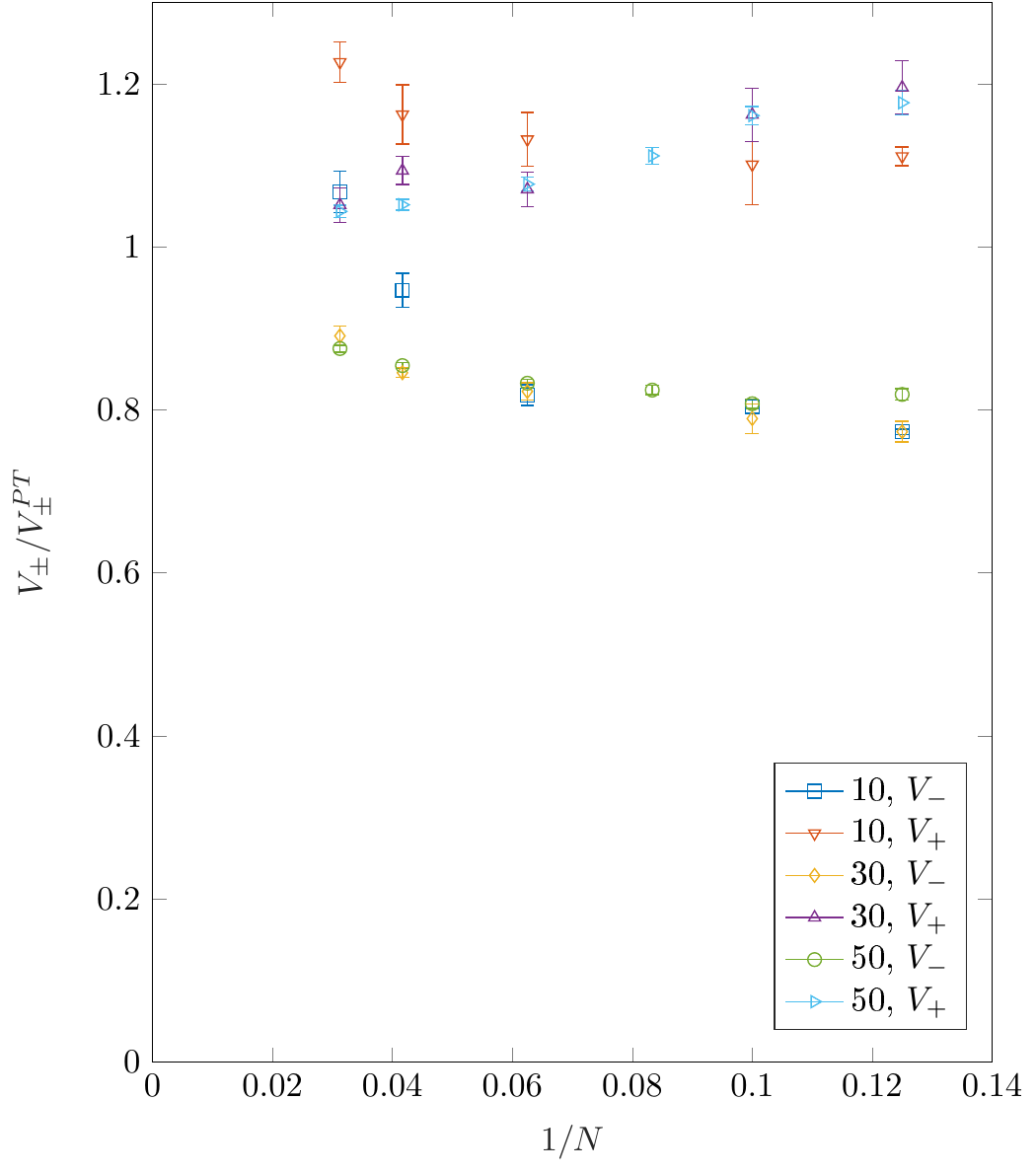}        
                %  \caption{ } 
        % \hspace{0.1cm}         
            \caption{The ratio of the exponential decays obtained from the MC measurements (via the fit \eqref{mf-fit}) and  and the PT prediction \eqref{mf-pt} for $g=10,30,50$ and various values of $N$.}
             \label{fig:VpmMC_VpmPT}
\end{figure}

As shown in Fig.~\ref{fig:fermioncorrelators}, a linear ($\sim N$) divergence and a strong dependence on the coupling $g$ appears in the measured ``masses'' $V_+$ and $V_-$ above.
%\colb{for both correlators}. 
%\colr{[with a sign of the divergence which is opposite for the $\theta$- and $\eta$- fermions, respectively] CAN WE FORGET THIS?]} 
%
A natural  guess is to relate this divergence to the $U(1)$ symmetry-breaking of our discretization, considering this as the fermionic counterpart of the bosonic effect $\langle x\rangle \neq 0$ which is also linearly divergent  -- see Section~\ref{sec:correlator} and discussion below \eqref{G_x}.  
In fact,  we may perform even in the continuum the simple exercise of evaluating these correlators on a vacuum with $\langle x\rangle \neq 0$.
Then at tree level the diagonal fermionic correlators read 
       \begin{eqnarray}\label{thetabroken}
\!\!\!\!\!\!\!\!\!\!\!\!\!\!\!\!\!\!\!\!\!\!\!\!
{C}_{\theta\theta}(t)_{\langle x\rangle\neq0} &\sim& \textstyle \frac{1}{2}\Big(1+\frac{2\,\left|\partial_{s}\langle x\rangle-m\frac{\langle x\rangle}{2}\right|}{\sqrt{4\,\left|\partial_{s}\langle x\rangle-m\frac{\langle x\rangle}{2}\right|^{2}+m^2}}\Big)\,e^{-t\,\widetilde{V_-}}+\frac{1}{2}\Big(1-\frac{2\,\left|\partial_{s}\langle x\rangle-m\frac{\langle x\rangle}{2}\right|}{\sqrt{4\,\left|\partial_{s}\langle x\rangle-m\frac{\langle x\rangle}{2}\right|^{2}+m^2}}\Big)\,e^{-t\,\widetilde{V_+}} \\\label{etabroken}
\!\!\!\!\!\!\!\!\!\!\!\!\!\!\!\!\!\!\!\!\!\!\!\!
{C}_{\eta\eta}(t)_{\langle x\rangle\neq0} &\sim&  \textstyle \frac{1}{2}\Big(1-\frac{2\,\left|\partial_{s}\langle x\rangle-m\frac{\langle x\rangle}{2}\right|}{\sqrt{4\,\left|\partial_{s}\langle x\rangle-m\frac{\langle x\rangle}{2}\right|^{2}+m^2}}\Big)\,e^{-t\,\widetilde{V_-}}
%\\
%&\qquad 
+\frac{1}{2}\Big(1+\frac{2\,\left|\partial_{s}\langle x\rangle-m\frac{\langle x\rangle}{2}\right|}{\sqrt{4\,\left|\partial_{s}\langle x\rangle-m\frac{\langle x\rangle}{2}\right|^{2}+m^2}}\Big)\,e^{-t\,\widetilde{V_+}}
      \end{eqnarray}
      with
\begin{align}\label{Vpm}
\widetilde{V}_{\pm} & =\sqrt{\frac{m^{2}}{4}+2\left|\partial_{s}\langle x\rangle-m\frac{\langle x\rangle}{2}\right|^{2}\pm2\left|\partial_{s}\langle x\rangle-m\frac{\langle x\rangle}{2}\right|\sqrt{\left|\partial_{s}\langle x\rangle-m\frac{\langle x\rangle}{2}\right|^{2}+\frac{m^{2}}{4}}}\,.
\end{align}
Clearly, as $\langle x\rangle=0$, it is $\widetilde{V}_{+}=\widetilde{V}_{-}\equiv m/2$ as it should~\footnote{It is worth emphasizing that the continuum theory has full $SO(6)\times U(1)$ symmetry, in particular $\langle x\rangle=0$. Namely, equations~\eqref{thetabroken},\eqref{etabroken} are written for illustrative purposes, supporting the interpretation that the divergence of the fermionic masses originates from symmetry breaking.}. Also, the sum of the correlators above reads
\be
C_\text{sum}(t)_{\langle x\rangle\neq0}= \frac{(C_{\theta\theta}(t)_{\langle x\rangle\neq0}+C_{\eta\eta}(t)_{\langle x\rangle\neq0})}{2}\sim e^{-t\,\widetilde{V_-}}+e^{-t\,\widetilde{V_+}}
\ee 
and thus justifies the choice for the fit functions in~\eqref{mf-fit}. 
We may also substitute in~\eqref{Vpm} the leading value for $\langle x\rangle$ obtained in perturbation theory in~\eqref{vevx} (considering $\partial_s\langle x\rangle=0$), thus obtaining for the exponential decay of the fermionic two-point functions above the expression
\be\label{mf-pt}
V_\pm^\text{PT}=\frac{m}{2}\frac{N\sqrt{2}}{g\,L\,m}\,\Big(\,\sqrt{1+\frac{(g\,L\,m)^2}{2\,N^2}}\pm 1\Big)~.
\ee
%Namely, a non-vanishing value of $\langle x\rangle$ would cause   the fermionic masses to deviate from the $m/2$ in a way which is directly driven by $\langle x\rangle$  and has opposite behavior for the two kinds of fermions. 
Plotting the exponential decays $V_\pm$ obtained via MC measurements against $V_\pm^\text{PT}$ as in Figure \ref{fig:VpmMC_VpmPT} one may notice a good convergence of the extrapolations to the expected values, at large $g$.

The observed divergence in the fermionic masses signals  that the continuum limit should be redefined.
In analogy with the case of chiral symmetry breaking of fermionic discretizations in lattice QCD (see e.g.~\cite{montvay}), one may interpret the divergence as an additive mass renormalisation of the bare coupling $m$ and proceed by 
studying the violation of the continuum Ward identities  on the lattice. We hope to report soon on this.

% As one does in QCD for the chiral symmetry, one can study the violation of the continuum Ward identities on the lattice and explicitly check that these violations vanish in the continuum limit. This program will be carried out analytically in lattice perturbation theory, and numerically in the non-perturbative regime. In parallel, we plan to investigate discretizations of the fermionic action, inspired to Ginsparg-Wilson fermions, which may preserve a larger symmetry group on the lattice. The procedure proposed in [VF1] to take the continuum limit is based on the assumption that the theory is super-renormalizable. Numerical simulations seem to confirm this idea, however this is a nontrivial statement as the action is non-polynomial in the fields. We plan to investigate more general ways to define lines of constant physics in the space of bare parameters. In principle this

\subsection{Impact of reweighting on observables}
\label{sec:reweighting}

As explained in section \ref{sec:simulations_strong}, we perform simulations with a fermionic operator \eqref{twistedmass} modified both via the replacement \eqref{fermionsintegration} with the absolute value of its Pfaffian and by a small twisted-mass term to avoid the instabilities due to its near-zero modes. 
The sign of the Pfaffian and the low modes of $O_F$ are then taken into account respectively by the  reweighting $W_s$ and $W_\mu$ in \eqref{reweight_factors}.  Here we comment on the impact of such reweighting on the observables.
%, which are computed according to \eqref{reweight}-\eqref{reweight_factors} in the modified model.

%IMP: Figure left shows less than 10 per cent of configurations have a minus sign, so the vev is not zero, while right  is bad.!
\begin{figure}[H]
%   \centering
           \includegraphics[scale=0.72]{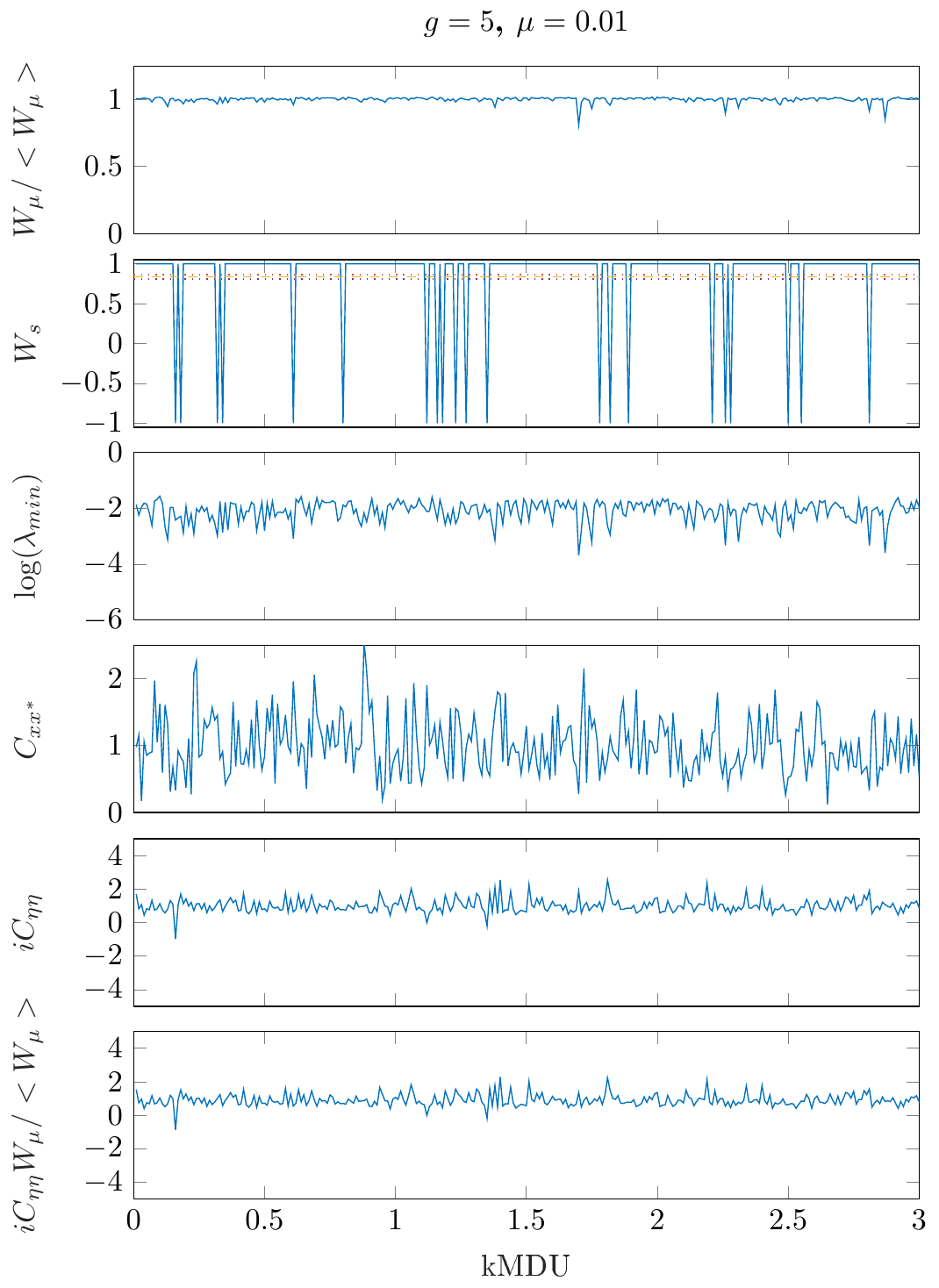}
               \hspace{0.1cm}
                \includegraphics[scale=0.72]{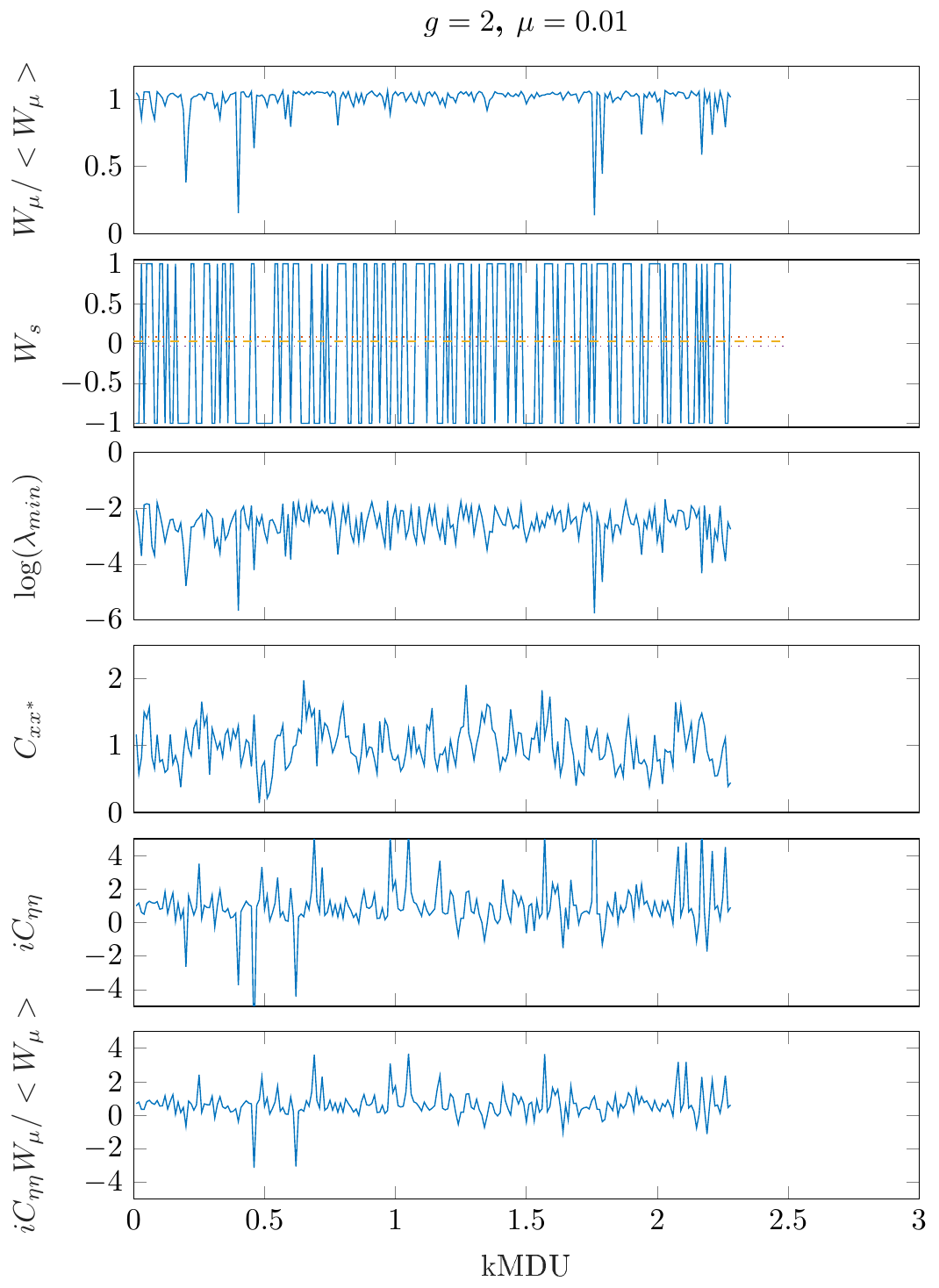}        
            \caption{Time history of the reweighting factors  $W_\mu$ and  $W_s$ in \eqref{reweight_factors}, the bosonic correlator $C_{xx}(t)$  and the fermionic correlator $C_{\eta\eta}(t)$ on two ensembles with $L=8$, $\mu=0.01$ and  $g=5$ (left), $g=2$ (right). The correlators are evaluated on a time-slice $t = T /4$. The last three lines are normalized, so that they average to 1 (e.g. the third line is actually $C_{xx}/\langle C_{xx}\rangle$). 
            For $g=2$ there  is a clear ``correlation'' between spikes in $W_\mu$ and the fermionic correlator.
 }
             \label{fig:MChistories}
\end{figure}

A pictorial way to study these effects is to look at the individual MC histories~\footnote{In MC simulations, vacuum expectation values are replaced by ensemble averages. Ensembles are generated by a Markov process (here, the RHMC) and the MC history is the change of the observable along the Markov process. In this sense it only makes sense to compare MC histories from the same simulation (see e.g.~\cite{montvay}).} of observables and reweighting factors, as well as the MC histories of their product (so, look at the observables ``before'' and ``after'' the reweighting).  
%In particular, here we can investigate the possible correlation between the reweighting factors in the vacuum expectation value~\eqref{reweight}-\eqref{reweight_factors}. 
Figure \ref{fig:MChistories} shows the MC evolution of the reweighting factors and of the observables as the simulation evolves, for two different values of the coupling $g=5$ (left) and $g=2$ (right) and the same value of the twisted-mass parameter $\mu=0.01$. There appear to be no (statistical)  correlation between the sign-reweighting $W_s$ and the observables, nor between $W_s$ and the $\mu$-reweighting $W_\mu$. However, as discussed in the previous section, small eigenvalues (and thus zero-crossings) are more probable to occur at lower $g$, which obviously reflects in a more severe sign problem (right diagram, $g=2$). 

As expected for bosonic observables, the fluctuations of the bosonic correlator are little correlated to those of the $\mu$-reweighting factor $W_\mu$. 
This is not so for the fermionic correlator. It is easy to spot a simultaneous occurrence of the negative peaks for the $\mu$-reweighting for $g=2$, upper right-diagram in Fig.\ref{fig:MChistories}, and the valleys in the value of the fermionic correlator (near MDU 20, 40 and 46).

\begin{figure}[H]
%   \centering         
     \includegraphics[scale=0.72]{L8_g5_Lm4_id8_rew-1.pdf}
         \hspace{0.5cm}         
           \includegraphics[scale=0.72]{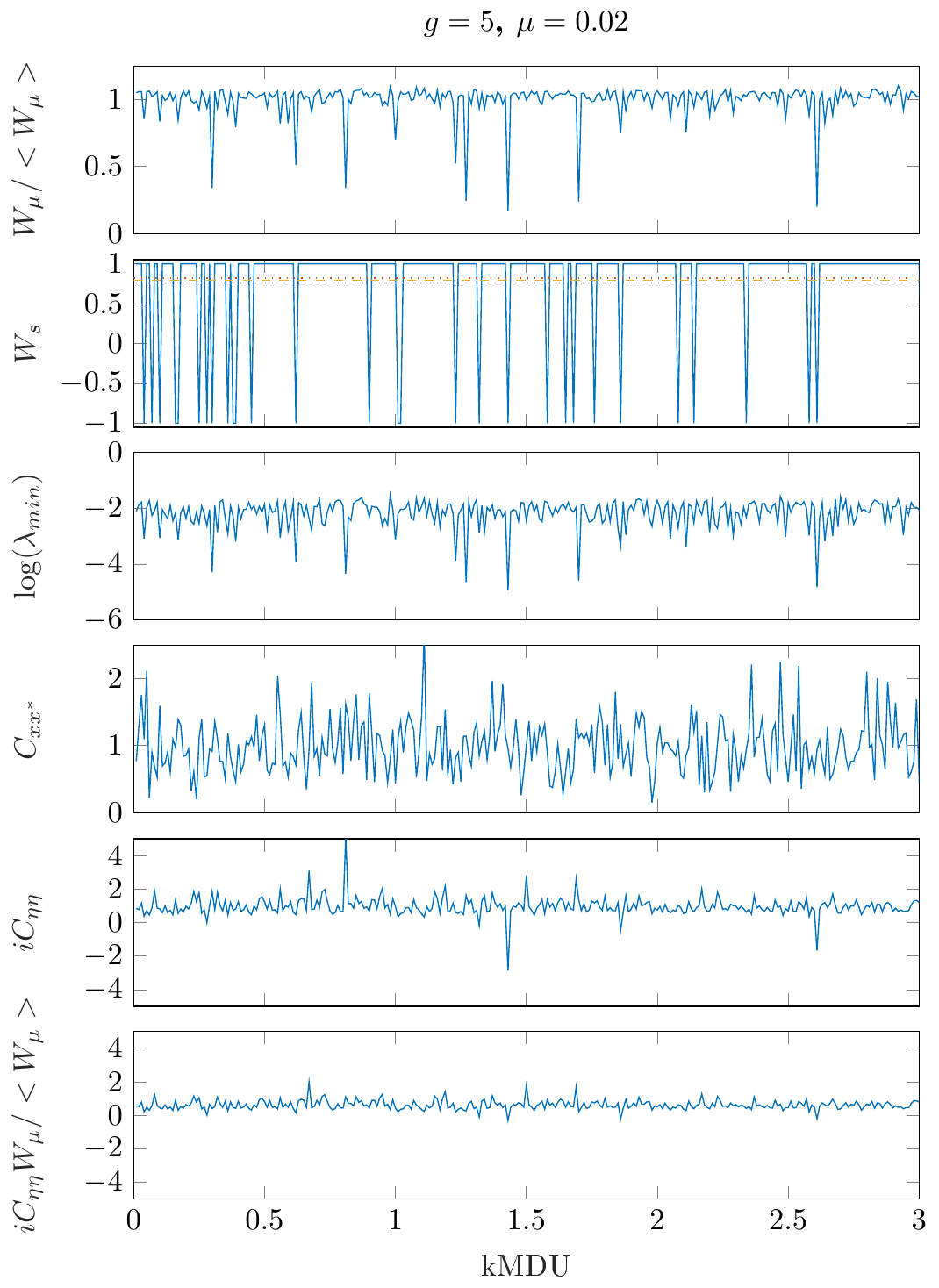}
       %    \hspace{0.1cm}
            \caption{ Time history of the reweighting factors  $W_\mu$ and  $W_s$ in \eqref{reweight_factors}, the bosonic correlator $C_{xx}(t)$  and the fermionic correlator $C_{\eta\eta}(t)$ on two ensembles with $L=8$, $g=5$ with two different values of the reweighting parameter, $\mu=0.01$ (left) and  $\mu=0.02$ (right). The correlators are evaluated on a time-slice $t = T /4$. The last three lines are normalized, so that they average to 1 (e.g. the third line is actually $C_{xx}/\langle C_{xx}\rangle$). For larger $\mu$, zero eigenvalues are more accessible and the fermionic correlator develops  spikes. The latter are cancelled after reweighting (sixth line). 
 }
             \label{fig:MChistories2}
\end{figure}

This correspondence between $W_\mu$ and the fermionic correlator is due to the sensitivity of the two-point function, built out of the inverse  fermionic operator, on the small eigenvalues of such operator,  to which $W_\mu$ is also (by definition) sensitive.  

In general, for the reweighting to work in practice, the fluctuations of the reweighting factor should be reasonably small (not to dominate the statistical error of the measured observable)~\cite{Luscher:2008tw, Bruno:2014lra,Bruno:2014jqa}. Such fluctuations clearly depend on the choice of $\mu$. 
%The optimal value of the twisted-mass parameter should mediate between two beneficial, but naturally exclusive, effects  (see e.g. \cite{Bruno:2014lra,Bruno:2014jqa}). On one side, 
A finite value of $\mu$ increases the ergodicity of the algorithm: field configurations with small eigenvalues of the original operator become statistically more significant in the path integral. 
%when it is the square root of the modified determinant to appear in the Boltzmann weight. 
On the other side, if $\mu$ becomes too large, the MC histories of fermionic correlators, which are controlled by the inverse of the modified operator, tend to develop sudden fluctuations.  These fluctuations are unphysical,  however they are cancelled in the ensemble average \eqref{reweight} by a smaller $W_\mu$.

That the choice of $\mu$ should be made with care is clear from Fig. \ref{fig:MChistories2}, where Monte Carlo histories are shown for two different values, $\mu=0.01$ (left) and $\mu=0.02$ (right), of the twisted-mass parameter and the same value $g=5$ of the coupling. A doubled value of $\mu$ enhances of a factor of 10 the fluctuations of the reweighting factor $W_\mu$ (first line). The sign-reweighting $W_s$ (second line, in which the red dotted lines  represent the average) also appears to be sensitive to the fact that zero eigenvalues are more accessible for larger $\mu$, something visible in the third line, where the logarithm of the lowest eigenvalue in the spectrum of $O_F O_F^\dagger$ appears.  
The bosonic correlator (fourth line) is as expected independent on the choice of the twisted-mass regulator. The situation is different for the fermionic correlator, which for larger $\mu$ develops  spikes (fifth line). The spikes are cancelled, as expected, after reweighting (sixth line). 
 
%the smallest the eigenvalue, the smallest the $\mu$-reweighting

%IMP:namely we measure them on the configurations with a fixed time.  
%IMP:namely we measure them on the configurations after a fixed distance in the fictitious MC time

%\colr{Namely, in an actual simulations formula ?? becomes the ensamble average 
%\be
%\langle \mathcal{O}\,w \,\rangle \rightarrow \sum_i\,\mathcal{O}_i\,w_i
%\ee
%}
%where i labels the configurations, or 
%\colr{In Fig \ref{MChistories} we 
% show the evolution of the reweighting factors and of the observables as the simulation proceeds,
%namely we measure them on the configurations after a fixed distance in the fictitious MC time }.

%This is explained  recalling  that at small quark masses both receive significant contributions from the smallest (in
%magnitude) eigenmodes of the fermionic operator. It is precisely this region where the reweighting term has the largest effect.
%To illustrate the cancellation between the fluctuations in W and fPP(x0), Figure \ref{MChistories} displays the time series of the two (top and central panel) at x0 = (T + a)/2 together with the product WfPP(x0); see Eq. (6.7) for its definition. Data for C101 and two values of ?0 is shown. As we can see, the larger ?0 leads to larger fluctuations in W and fPP(x0), as expected. In the product, however, they cancel and the average value ?WfPP(x0)?/?W? is then consistent within the statistical errors between the two ensembles.

%1h:45
A more quantitative way to see the effect of reweighting on the observables is a study of the covariance  between the observables $\mathcal{O}$ and the reweighting factors $W$~\footnote{In particular, a vanishing covariance (from which  $\langle\mathcal{O}\,W\rangle = \langle\mathcal{O}\rangle\, \langle W\rangle$) would imply the cancellation of $\langle W\rangle$ in~\eqref{reweight}. In this case the reweighting would not change the value of the observable, but only its variance.}. While we have observed that, as expected, the largest covariance is between the $\mu$-reweighting and the value of the lowest eigenvalue of the fermionic operator, we could not in general draw a conclusive picture  from this study because the effects are smaller than the statistical error.  

%Yet another way to observe the effect of reweighting is to  look at 

Table \ref{t:table2} shows the effect of reweighting on the numerical values of the ensemble averages at one value of the coupling ($g=5$) and two values $\mu=0.01, 0.02$ of the $\mu$-reweighting.
\begin{table}[H]
\begin{longtable}[c]{@{}lll@{}}
\toprule\addlinespace
& $g=5$, $\mu=0.01$ & $g=5$, $\mu=0.02$
\\\addlinespace
\midrule\endhead
$<C_{xx^*}>$ & 0.1620(44) & 0.1619(31)
\\\addlinespace
$<C_{xx^*}>_{W_s}$ & 0.1620(44) & 0.1624(31)
\\\addlinespace
$<C_{xx^*}>_{W}$ & 0.1604(49) & 0.1643(38)
\\\addlinespace
$<C_{\eta\eta^*}>$ & 0.1464(32) & 0.1502(40)
\\\addlinespace
$<C_{\eta\eta^*}>_{W_s}$ & 0.1461(32) & 0.1505(34)
\\\addlinespace
$<C_{\eta\eta^*}>_{W}$ & 0.1508(37) & 0.1584(42)
\\\addlinespace
\bottomrule
\caption{Effect of the reweighting on the two-point functions.}
\label{t:table2}
\end{longtable}
\end{table}
It is interesting to notice that the sign-reweighting  seems practically not to have effect on the measured observables.  
About the $\mu$-reweighting, although not statistically significant, the effect is larger for the fermionic correlator.  
%the Table confirms that only the fermionic correlator is influenced by the choice of the twisted-mass regulator.

Our last observation is about the behavior of the reweighting factors with the lattice spacing. 
\begin{figure}[H]
   \centering
 \includegraphics[scale=0.8]{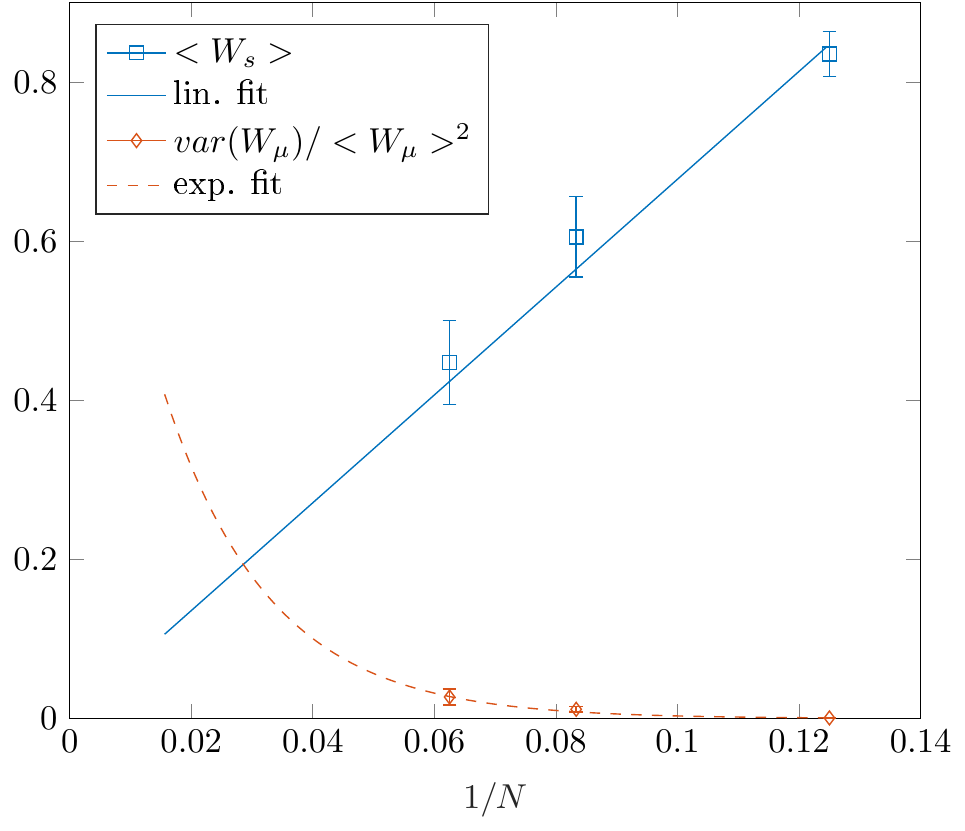}
 \caption{Lattice spacing dependence of $<W_s>$ and
variance of $W_\mu$ at $g=5$ and $\mu=0.01$. }
\label{fig:volume_dep}
\end{figure}
This is done in Fig. \ref{fig:volume_dep}. 
The sign-reweighting  $W_s$  shows a moderate (linear) dependence and tends towards
zero for $1/N\to 0$. However, in the region of our simulations it is well above zero. The fluctuations of the $\mu$-reweighting (at fixed $\mu$) are small and compatible with an exponential dependence on $N$.
%which is expected for a 
%reweighting by ratios of \colb{fermionic determinants which scale exponentially with the volume of the lattice}. 
Extrapolating these points simulations up to $N\sim 32$ seems feasible at $g=5$.

%\section{The new observable}
%
%\colb{Derivative with respect to $m$  in Nils' thesis, or in the Appendix?}.

\section*{Acknowledgements}

We are particularly grateful to Radu Roiban for several discussions. We thank Luigi Del Debbio, Michele Della Morte, Agostino Patella, Rainer Sommer and the members of the Innovative Training Network EuroPLEx for discussions. 
The research of LB received funding from the European Union's Horizon 2020 research and innovation programme under the Marie Sklodowska-Curie grant agreement No 749909.
The research of VF received funding from the STFC grant ST/S005803/1, from the Einstein Foundation Berlin through an Einstein Junior Fellowship, and was  supported in part by Perimeter Institute for Theoretical Physics and the Simons Foundation through a Simons Emmy Noether Fellowship. 
The research of EV received funding by the FAPESP grants 2014/18634-9 and 2016/09266-1, and by the STFC grant ST/P000762/1.

 %%%%%%%%%%%%%%%%%%%%%%%%%%%%%%%%%%%%%
 %%%%%%%%%%%%%%%%%%%%%%%%%%%%%%%%%%%%%

\appendix

\section{Conventions and matrix algebra}
\label{app:continuum}
In the action \eqref{cuspaction} we used the  six $4\times 4$ matrices $(\rho^M)_{ij}$, off-diagonal blocks of the $SO(6)$, $8\times 8$  Dirac matrices
in chiral representation
\begin{equation} 
\gamma^M\equiv \begin{pmatrix}
0  & \rho^\dagger_M   \\
 \rho^M   &  0 
\end{pmatrix}
=
\begin{pmatrix}
0  & (\rho^M)^{ij}   \\
(\rho^M)_{ij}   &  0 
\end{pmatrix}
\end{equation}
for which 
\begin{align}\label{relationsrho}
\rho_{ij}^M &=- \rho_{ji}^M\,, &
   (\rho^M)^{il}\rho_{lj}^N + (\rho^N)^{il}\rho_{lj}^M
 &=2\delta^{MN}\delta_j^i\,, &
  (\rho^M)^{ij}&\equiv  -
%+ in original non-MTT conventions
 (\rho_{ij}^{M})^* \,.
 %=(\rho^M{}^\dagger)^{ij}\ ,~~~~~~~~ 
 \end{align}
%\colb{ The two off-diagonal blocks, carrying upper and lower indices respectively, are related by $(\rho^M)^{ij}=-(\rho^M_{ij})^*\equiv(\rho^M_{ji})^*$, so that indeed the block with upper indices is the conjugate transpose $(\rho_{M}^{\dagger})^{ij}$  of the block with lower indices.}
A possible explicit  representation is
\begin{align}
\rho^1_{ij}&=\left(\begin{matrix}0&1&0&0\\-1&0&0&0\\0&0&0&1\\0&0&-1&0
\end{matrix}\right)\,,&
\rho^2_{ij}&=\left(\begin{matrix}0&\mathrm{i}&0&0\\-\mathrm{i}&0&0&0\\0&0&0&-\mathrm{i}\\0&0&\mathrm{i}&0
\end{matrix}\right)\,,&
\rho^3_{ij}&=\left(\begin{matrix}0&0&0&1\\0&0&1&0\\0&-1&0&0\\-1&0&0&0
\end{matrix}\right)\,,\nonumber \\
\rho^4_{ij}&=\left(\begin{matrix}0&0&0&-\mathrm{i}\\0&0&\mathrm{i}&0\\0&-\mathrm{i}&0&0\\ \mathrm{i}&0&0&0
\end{matrix}\right)\,,&
\rho^5_{ij}&=\left(\begin{matrix}0&0&\mathrm{i}&0\\0&0&0&\mathrm{i}\\-\mathrm{i}&0&0&0\\0&-\mathrm{i}&0&0
\end{matrix}\right)\,,&
\rho^6_{ij}&=\left(\begin{matrix}0&0&1&0\\0&0&0&-1\\-1&0&0&0\\0&1&0&0
\end{matrix}\right)\,.\nonumber
\end{align}
The $SO(6)$ generators are built out of the $\rho$-matrices via
\begin{equation}
 \rho^{MN}{}^i{}_{ j} \equiv\frac{1}{2}[  (\rho^M)^{il}\rho_{lj}^N
- (\rho^N)^{il}\rho_{lj}^M ]
\end{equation}
and the following identities hold
\begin{align}\label{rhomatrices}
  (\rho^{MN})^i_{\hphantom{i}j}&=\left( (\rho^{MN})_i^{\hphantom{i}j}\right)^* &  (\rho^{MN})^i_{\hphantom{i}j}&=-(\rho^{MN})_j^{\hphantom{j}i} \,,
\end{align}
where in the last equation we used that $\frac{1}{2}({\rho^M}^{i\ell}\,\rho^N_{\ell j} -{\rho^N}^{i\ell}\,\rho^M_{\ell j} )=-\frac{1}{2}(\rho^M_{j\ell}\,{\rho^N}^{\ell i} -\rho^N_{j\ell}\,{\rho^M}^{\ell i} )$. Useful flipping rules are
\begin{eqnarray}
\eta\,\rho^M\,\theta&=&\eta^i\,\rho^M_{ij}\,\theta^j=-\theta^j\,\rho^M_{ij}\,\eta^i=\theta^j\,\rho^M_{ji}\,\eta^i\equiv \theta^i\,\rho^M_{ij}\,\eta^j=\theta\,\rho^M\,\eta\\
\eta^\dagger\rho^\dagger_M\,\theta^\dagger&=&\eta_i\,{\rho^M}^{ij}\,\theta_j=-\theta_j\,{\rho^M}^{ij}\,\eta_i=\theta_j\,{\rho^M}^{ji}\,\eta_i\equiv \theta_i\,{\rho^M}^{ij}\,\eta_j=\theta^\dagger \rho^\dagger_M\,\eta^\dagger\\
\eta_i\,(\rho^{MN})^i_{\hphantom{i}j}\,\theta^j&=&-\theta^j\,(\rho^{MN})^i_{\hphantom{i}j}\,\eta_i=\theta^j\,(\rho^{MN})_j^{\hphantom{j}i}\,\eta_i\equiv\theta^i\,(\rho^{MN})_i^{\hphantom{i}j}\,\eta_j~.
\end{eqnarray}

In the main text, for the steps leading from \eqref{eq:quarticaction} to \eqref{newquarticaction} we used the following additional properties
\begin{align}
 (\rho^M)^{im} (\rho^M)^{kn}&=2\e^{imkn} \\
 (\rho^M)^{im} (\rho^M)_{nj}&=2\left(\d^i_j \d^m_n -\d^i_n \d^m_j\right)\\
 \e^{imkn} (\rho^M)_{mj}(\rho^L)_{nl}+\e_{mjnl}(\rho^M)^{im}(\rho^L)^{kn}&=(\rho^{\{M})^{ik}(\rho^{L\}})^{jl}+\d^k_j(\rho^L)^{im}(\rho^M)_{ml}+\d^i_l(\rho^M)^{km}(\rho^L)_{mj}\nonumber\\
 &+\d^{ML}\left(-4\d^i_l \d^k_j+2\d^i_j\d^k_l\right)\\
 -{(\rho^{MN})^i}_j {(\rho^{ML})^k}_l n_N n_L&=-2(\rho^{N})^{ik} (\rho^{L})_{jl} n_N n_L-\d^i_j \d^k_l+2\d^i_l \d^k_j
\end{align}
leading to the identification
\begin{align}
 \left(i\, \eta_i {(\rho^{MN})^i}_j n^N \eta^j\right)^2=-3 (\eta^2)^2+2\eta_i (\rho^N)^{ik} n_N \eta_k \eta^j (\rho^L)_{jl} n_L \eta^l
\end{align}

Around equation \eqref{sigmatilde} we also defined
\begin{align}\label{rel2}
 \S_i^j=\eta_i \eta^j \qquad \tilde{\S}_j^i=(\rho^N)^{ik}n_N (\rho^L)_{jl}n_L \eta_k \eta^l
\end{align}
where we simply indicate $\S_j^i={\S^i}_j={\S_j}^i $ since
\begin{align}
 {\S^i}_j\equiv ({\S_i}^j)^\ast=(\eta^j)^\ast(\eta_i)^\ast=\eta_j \eta^i={\S_j}^i 
\end{align}
and similarly for $\tilde \S$. It is simple to check that
\begin{align}\label{rel3}
 \S_i^j \S^i_j&=-(\eta^2)^2 & \tilde\S_i^j \tilde\S^i_j&=-(\eta^2)^2 & \S^i_j \tilde\S^j_i&=-\left|\eta_i (\rho^N)^{ik} n_N \eta_k\right|^2\\
 (\S_i^j)^\ast &=\S_j^i &  (\tilde \S_i^j)^\ast &=\tilde \S_j^i 
\end{align}

We conclude this section with a detailed counting of the degrees of freedom implied in the Hubbard Stratonovich transformation \eqref{HubbardStratonovich}. The $4\times 4$  matrix $\S_+$ is hermitian and contains 16 real d.o.f. One can project the two indices $i$ and $j$ onto irreducible $su(4)$ representations
\begin{equation}
 \mathbf{4}\otimes \bar{\mathbf{4}}=\mathbf{15}\oplus \mathbf{1}
\end{equation}
or, more explicitly
\begin{equation}
 {\S_+}_i^j=\frac14 {(\rho^{MN})^j}_i S_{MN}+\frac12\d_i^j S
\end{equation}
The term $\Tr\S_+\S_+$ in the Lagrangian would read
\begin{equation}
 \Tr\S_+\S_+=\frac12 S_{MN}S^{MN} +S^2
\end{equation}
This is a sum of $15+1$ real terms (remember $S_{MN}$ is an antisymmetric $6\times6$ matrix). To any of these terms one can associate, via a Hubbard Stratonovich transformation, a real scalar field (therefore 15 scalars $\phi_{MN}$ in the adjoint and one in the singlet). Then, by the opposite procedure one can rebuild the matrix $\phi_i^j$ used in \eqref{HubbardStratonovich}. This proves that the matrix $\phi_i^j$ is hermitian.
 
 \section{One-point function for $x$, $x^*$}
 \label{app:onepoint}
 
In the continuum, the action \eqref{S_cusp} and its  linearized version (\ref{Scuspquadratic}), (\ref{OF}), (\ref{A}) enjoy the $SO(6)\times U(1)$ symmetry of the cusp background. In particular, the $U(1)$ invariance implies $\braket{x}=\braket{x^*}=0$. The Wilson-like discretization \eqref{OFgen}-\eqref{Wilsonshiftgen}  adopted in this paper for the fermionic sector breaks the $U(1)$ symmetry, and as a consequence the fields $x$, $x^*$ acquire then a non-trivial, in fact divergent, 1-point function. We evaluate here this one-point function at leading order, $\mathcal{O}(g^{-1})$, in lattice perturbation theory.  

The continuum sigma-model loop expansion for this model (in AdS light-cone gauge) is studied in~\cite{Giombi,Giombi:2010bj}. % (and reviewed for example in~\cite{Bianchi:2016yfl,Vescovi:2016zzu}
%~\footnote{The seminal references \cite{Gubser:2002tv,Frolov:2002av}, evaluating the leading and subleading corrections to the scaling function,  use a different gauge-fixing and  therefore different Lagrangian.})
, and a first calculation in lattice perturbation theory appears in Section 3 (see also Appendix A) of~\cite{Bianchi:2016cyv}. Here we recall that in order to perform a perturbative computation, in the continuum and on the lattice, one cannot simply expand around the trivial vacuum where all the fields are set to zero -- this is prevented by the presence of inverse powers of the radial coordinate $z$ in the Lagrangian. One proceeds then picking one of the degenerate ``null cusp'' vacua corresponding to the $SO(6)$ directions of $z^M$ (this breaks the  $SO(6)$ symmetry to a $SO(5)$),   say $u^M=(0,\,0,\,0,\,0,\,0,\,1)$, where $u^M$, with $u^M u^M=1$ are part of the standard definition of Poincare' patch coordinates $ {\tilde z}^M = e^{\tilde \phi} \tilde u^M$,  $ {\tilde z}= e^{\tilde \phi}$. 
In terms of
\be
{\tilde u}{}^{a}=  \frac{y^{a}}{1+\frac{1}{4}y^2}~, \ \ \ \ 
{\tilde u}{}^{6} =  \frac{1-\frac{1}{4}y^2}{1+\frac{1}{4}y^2}  \ , \ \ \ \ \ \ \ \ \
~~~~ y^2\equiv \sum_{a=1}^5 (y^a)^2\ , \ \ \ \ \ a=1,...,5 \ , \label{exp6} 
\ee
the  vacuum corresponds then to $y^a =\phi=0$.  \\
Because of our Wilson discretization, the diagonal fermionic propagators $C_{\eta^i\eta^i}$ and $C_{\eta_i\eta_i}$, corresponding to the two lower diagonal entries of \eqref{Kinverse}, are non-vanishing.  
%(\colb{in this context, they are called Wilson fermions}). 
The  cubic interaction
%~\footnote{One reads this interaction term from the two lower diagonal entries of \eqref{OF}, and considering that  $u^M$ are part of the redefinition~\cite{Giombi} of $z^M= e^{\phi}u^M$, where $\phi$ is the radial coordinate  $z=e^{\phi}$, and the vacuum is  }
\be\label{Scubic}
S_{x\eta\eta}=2g\,\int dt\,ds \Big[ \eta^i \,\rho_{ij}^M\,\eta^j\,(\partial_s x-\textstyle\frac{m}{2}x)\,u^M-\eta_i\,\rho^{ij}_M \,\eta_j \,(\partial_s x^*-\textstyle\frac{m}{2}x^*)\,u^M\Big]\,,
\ee 
gives then a contribution at order $1/g$   to the 1-point function of $x, x^*$ through a tadpole graph with a single fermionic loop.  
In momentum space the relevant propagators read 
\begin{eqnarray}\label{propagator_bos}
C_{x x^*} (p_0,p_1)&=&\frac{1}{g}\frac{2}{\hat{p}^2+\frac{m^2}{2}}\\\label{propagator_ferm_eta_low}
C_{\eta_i\eta_j}(p_0,q_1)&=&\frac{a}{g}\,[K_F^{-1} (p_0, p_1)]_{44} =-\frac{a\,r}{2\,g}\frac{(\hat{p}_0^2-i \,\hat{p}_1^2)\,\rho^M_{ij}u^M}{\left[\textrm{det}{K}_{F}(p_0,p_1) \right]^{1/8}}\\ \label{propagator_ferm_eta_high} 
C_{\eta^i\eta^j}(p_0,q_1)&=&\frac{a}{g}\,[K_F^{-1} (p_0, p_1)]_{33} =\frac{a\,r}{2\,g}\frac{(\hat{p}_0^2+i \,\hat{p}_1^2)\,\rho_M^{ij}u^M}{\left[\textrm{det}{K}_{F}(p_0,p_1) \right]^{1/8}}\,.
\end{eqnarray} 
where the bosonic one \eqref{propagator_bos} is obtained from the continuum~\cite{Giombi,Bianchi:2016cyv}  with the naive replacement $p_\mu\to \hat{p}_\mu$, and the fermionic  propagators are taken from \eqref{Kinverse}-\eqref{detKF}-\eqref{Kdagger}.

For the $x$-field, Wick-contracting and using \eqref{propagator_bos} and \eqref{propagator_ferm_eta_low} and the second term in \eqref{Scubic}, one writes  formally, in momentum space, at leading order (LO) in $1/g$ expansion
\be\!\!\!\!\!\!\!\!\!
\!\!\!\!\!\!\langle \tilde{x}(q)\rangle_\text{LO}=\frac{8\,r\,a}{g} \,\delta^{(2)}(q)\,\frac{i\,\hat{q}_1-\textstyle\frac{m}{2}}{\hat{q}^2+\frac{m^2}{2}}\,u^M \rho_M^{ij}\rho^N_{ij}u^N\,\iint_{-\frac{\pi}{a}}^{\frac{\pi}{a}}\!\frac{d^2p}{(2\pi)^2}\,
\frac{\hat{p}_0^2-i \,\hat{p}_1^2}{\mathring{p_0} ^{2}+\mathring{p_1} ^{2}+\frac{m^{2}}{4}+\frac{a^2 \,r^{2}}{4}\left(\hat{p}_{0}^{4}+\hat{p}_{1}^{4}\right)}\,,
\ee
where we denoted with $q$ the 2-momentum of the external bosonic field $x$, with $p_0,p_1$ the 2-momentum of the fermion in the loop and we used \eqref{detKF}.  Above, $\delta^{(2)}(q)$ is the momentum conservation at the vertex.
Rescaling the momenta with the lattice spacing, using that  \eqref{relationsrho} implies $ \rho_M^{ij}\rho^N_{ij}u^Mu^N=-4$ and setting $r=1$ one obtains 
\be
 \langle \tilde{x}(q)\rangle_\text{LO}=-\frac{32}{g\,a}\, (1-i)\,I(M)\,\delta^{(2)}(q)\frac{i\,\hat{q}_1-\textstyle\frac{m}{2}}{\hat{q}^2+\frac{m^2}{2}}\,,
\ee
where ($M=m\,a$)
\be
I(M)=\int_{-\pi}^{\pi}\frac{dp_0\,dp_1}{(2\pi)^2}\,\frac{\sin^2\frac{p_0}{2}}{\sin^2p_0+\sin^2p_1+4\sin^4\frac{p_0}{2}+4\sin^4\frac{p_1}{2}+M^2}\,,~~\text{with}~~ I(0)=\frac{1}{32}~.
\ee 
Fourier transforming back in position space one obtaines
\bal
\!\!\!\!\!
\langle x\rangle_\text{LO}& =  \iint_{-\frac{\pi}{a}}^{\frac{\pi}{a}}dq_0 dq_1 \,e^{-i t \,q_0-i s\,q_1}\, \langle \tilde{x}(q)\rangle  \\
& =-\frac{32}{g\,a}\, (1-i)\,I(M)\, \iint_{-\frac{\pi}{a}}^{\frac{\pi}{a}}dq_0 dq_1 \,\delta(q_0)\,\delta(q_1)\,e^{-i t \,q_0-i s\,q_1}\, \frac{\frac{i}{a}\,\sin\frac{q_1}{2}-\textstyle\frac{m}{2}}{\frac{1}{a^2}\sin^2\frac{q_0}{2}+\frac{1}{a^2}\sin^2\frac{q_1}{2}+\frac{m^2}{2}}\\
\!\!\!\!\!&=-\frac{32}{g}\, (1-i)\,I(M)\,\frac{1}{m\,a}~.
\eal
Using that in the continuum limit $a\to 0$ the product $m\,L=M N$ is fixed  and  that $I(0)=\frac{1}{32}$, we find that  the one-point function diverges linearly in $N$ ($=L/a$) as
\be\label{vevx}
\langle x\rangle_\text{LO} =\frac{N}{g\,m\,L} (1-i)\,.
\ee
This result  is perfectly consistent with the plot of  Fig.~\ref{fig:x-vev} for several values of (large) $g$.
Repeating the computation for the field $x^*$, therefore using the first term in \eqref{Scubic} and  \eqref{propagator_ferm_eta_low}, it is easy to verify that
\be\label{vevxstar}
\langle x^*\rangle_\text{LO}=\frac{N}{g\,m\,L} (1+i)\,.
\ee
The two equations above are consistent with \eqref{vevrvevim} at leading ($1/g$) order in sigma-model perturbation theory. 
 
\bibliographystyle{nb}
\bibliography{Ref_strings_lattice}

%bibliography generated by nb.bst v1.01 (C) 2003-2010 Niklas Beisert
\begin{thebibliography}{10}
\ifx\href\asklfhas\newcommand{\href}[2]{#2}\fi
\ifx\arxivref\asklfhas\newcommand{\arxivref}[2]{\href{http://arxiv.org/abs/#1}{#2}}\fi
\ifx\doiref\asklfhas\newcommand{\doiref}[2]{\href{http://dx.doi.org/#1}{#2}}\fi
\raggedright
\small
\parskip 0pt

\bibitem{Catterall_physrept}
S.~Catterall, D.~B.~Kaplan and M.~Unsal,
\textit{``{Exact lattice supersymmetry}''},
\textsf{\doiref{10.1016/j.physrep.2009.09.001}{Phys.~Rept.~484,~71~(2009)}},
\texttt{\arxivref{0903.4881}{arxiv:0903.4881}}.
%%CITATION = ARXIV:0903.4881;%%

\bibitem{Schaich:2015ppr}
D.~Schaich,
\textit{``{Aspects of lattice N=4 supersymmetric Yang--Mills}''},
\textsf{PoS~LATTICE2015,~242~(2015)},
\texttt{\arxivref{1512.01137}{arxiv:1512.01137}},
in: \textit{``{Proceedings, 33rd International Symposium on Lattice Field
  Theory (Lattice 2015)}''},
242p.
%%CITATION = ARXIV:1512.01137;%%

\bibitem{Joseph:2015xwa}
A.~Joseph,
\textit{``{Review of Lattice Supersymmetry and Gauge-Gravity Duality}''},
\textsf{\doiref{10.1142/S0217751X15300549}{Int.~J.~Mod.~Phys.~A30,~1530054~(2015)}},
\texttt{\arxivref{1509.01440}{arxiv:1509.01440}}.
%%CITATION = ARXIV:1509.01440;%%

\bibitem{Bergner:2016sbv}
G.~Bergner and S.~Catterall,
\textit{``{Supersymmetry on the lattice}''},
\texttt{\arxivref{1603.04478}{arxiv:1603.04478}}.
%%CITATION = ARXIV:1603.04478;%%

\bibitem{Schaich:2016jus}
D.~Schaich, S.~Catterall, P.~H.~Damgaard and J.~Giedt,
\textit{``{Latest results from lattice N=4 supersymmetric Yang--Mills}''},
\textsf{PoS~LATTICE2016,~221~(2016)},
\texttt{\arxivref{1611.06561}{arxiv:1611.06561}},
in: \textit{``{Proceedings, 34th International Symposium on Lattice Field
  Theory (Lattice 2016): Southampton, UK, July 24-30, 2016}''},
221p.
%%CITATION = ARXIV:1611.06561;%%

\bibitem{Berkowitz:2016jlq}
E.~Berkowitz, E.~Rinaldi, M.~Hanada, G.~Ishiki, S.~Shimasaki and P.~Vranas,
\textit{``{Precision lattice test of the gauge/gravity duality at
  large-$N$}''},
\textsf{\doiref{10.1103/PhysRevD.94.094501}{Phys.~Rev.~D94,~094501~(2016)}},
\texttt{\arxivref{1606.04951}{arxiv:1606.04951}}.
%%CITATION = ARXIV:1606.04951;%%

\bibitem{Rinaldi:2017mjl}
E.~Rinaldi, E.~Berkowitz, M.~Hanada, J.~Maltz and P.~Vranas,
\textit{``{Toward Holographic Reconstruction of Bulk Geometry from Lattice
  Simulations}''},
\textsf{\doiref{10.1007/JHEP02(2018)042}{JHEP~1802,~042~(2018)}},
\texttt{\arxivref{1709.01932}{arxiv:1709.01932}}.
%%CITATION = ARXIV:1709.01932;%%

\bibitem{Catterall:2017lub}
S.~Catterall, R.~G.~Jha, D.~Schaich and T.~Wiseman,
\textit{``{Testing holography using lattice super-Yang-Mills theory on a
  2-torus}''},
\textsf{\doiref{10.1103/PhysRevD.97.086020}{Phys.~Rev.~D97,~086020~(2018)}},
\texttt{\arxivref{1709.07025}{arxiv:1709.07025}}.
%%CITATION = ARXIV:1709.07025;%%

\bibitem{Schaich:2018mmv}
D.~Schaich,
\textit{``{Progress and prospects of lattice supersymmetry}''},
\textsf{\doiref{10.22323/1.334.0005}{PoS~LATTICE2018,~005~(2019)}},
\texttt{\arxivref{1810.09282}{arxiv:1810.09282}},
in: \textit{``{Proceedings, 36th International Symposium on Lattice Field
  Theory (Lattice 2018): East Lansing, MI, United States, July 22-28, 2018}''},
005p.
%%CITATION = ARXIV:1810.09282;%%

\bibitem{Roiban}
R.~McKeown and R.~Roiban,
\textit{``{The quantum $AdS_5 \times S^5$ superstring at finite coupling}''},
\texttt{\arxivref{1308.4875}{arxiv:1308.4875}}.
%%CITATION = ARXIV:1308.4875;%%

\bibitem{POS2015}
V.~Forini, L.~Bianchi, M.~S.~Bianchi, B.~Leder and E.~Vescovi,
\textit{``{Lattice and string worldsheet in AdS/CFT: a numerical study}''},
\textsf{PoS~LATTICE2015,~244~(2016)},
\texttt{\arxivref{1601.04670}{arxiv:1601.04670}},
in: \textit{``{Proceedings, 33rd International Symposium on Lattice Field
  Theory (Lattice 2015): Kobe, Japan, July 14-18, 2015}''},
244p.
%%CITATION = ARXIV:1601.04670;%%

\bibitem{Bianchi:2016cyv}
L.~Bianchi, M.~S.~Bianchi, V.~Forini, B.~Leder and E.~Vescovi,
\textit{``{Green-Schwarz superstring on the lattice}''},
\textsf{\doiref{10.1007/JHEP07(2016)014}{JHEP~1607,~014~(2016)}},
\texttt{\arxivref{1605.01726}{arxiv:1605.01726}}.
%%CITATION = ARXIV:1605.01726;%%

\bibitem{Forini:2017mpu}
V.~Forini, L.~Bianchi, B.~Leder, P.~Toepfer and E.~Vescovi,
\textit{``{Strings on the lattice and AdS/CFT}''},
\textsf{PoS~LATTICE2016,~206~(2016)},
\texttt{\arxivref{1702.02005}{arxiv:1702.02005}},
in: \textit{``{Proceedings, 34th International Symposium on Lattice Field
  Theory (Lattice 2016): Southampton, UK, July 24-30, 2016}''},
206p.
%%CITATION = ARXIV:1702.02005;%%

\bibitem{Forini:2017ene}
V.~Forini,
\textit{``{On regulating the AdS superstring}''},
\texttt{\arxivref{1712.10301}{arxiv:1712.10301}}.
%%CITATION = ARXIV:1712.10301;%%

\bibitem{BES}
N.~Beisert, B.~Eden and M.~Staudacher,
\textit{``{Transcendentality and Crossing}''},
\textsf{\doiref{10.1088/1742-5468/2007/01/P01021}{J.~Stat.~Mech.~0701,~P01021~(2007)}},
\texttt{\arxivref{hep-th/0610251}{hep-th/0610251}}.
%%CITATION = HEP-TH/0610251;%%

\bibitem{Basso:2013vsa}
B.~Basso, A.~Sever and P.~Vieira,
\textit{``{Spacetime and Flux Tube S-Matrices at Finite Coupling for N=4
  Supersymmetric Yang-Mills Theory}''},
\textsf{\doiref{10.1103/PhysRevLett.111.091602}{Phys.~Rev.~Lett.~111,~091602~(2013)}},
\texttt{\arxivref{1303.1396}{arxiv:1303.1396}}.
%%CITATION = ARXIV:1303.1396;%%

\bibitem{Basso:2013aha}
B.~Basso, A.~Sever and P.~Vieira,
\textit{``{Space-time S-matrix and Flux tube S-matrix II. Extracting and
  Matching Data}''},
\textsf{\doiref{10.1007/JHEP01(2014)008}{JHEP~1401,~008~(2014)}},
\texttt{\arxivref{1306.2058}{arxiv:1306.2058}}.
%%CITATION = ARXIV:1306.2058;%%

\bibitem{Fioravanti:2015dma}
D.~Fioravanti, S.~Piscaglia and M.~Rossi,
\textit{``{Asymptotic Bethe Ansatz on the GKP vacuum as a defect spin chain:
  scattering, particles and minimal area Wilson loops}''},
\textsf{\doiref{10.1016/j.nuclphysb.2015.07.007}{Nucl.~Phys.~B898,~301~(2015)}},
\texttt{\arxivref{1503.08795}{arxiv:1503.08795}}.
%%CITATION = ARXIV:1503.08795;%%

\bibitem{Bonini:2018mkg}
A.~Bonini, D.~Fioravanti, S.~Piscaglia and M.~Rossi,
\textit{``{Fermions and scalars in $\mathcal{N} = 4$ Wilson loops at strong
  coupling and beyond}''},
\textsf{\doiref{10.1016/j.nuclphysb.2019.114644}{Nucl.~Phys.~B,~114644~(2019)}},
\texttt{\arxivref{1807.09743}{arxiv:1807.09743}}.
%%CITATION = ARXIV:1807.09743;%%

\bibitem{MT2000}
R.~Metsaev and A.~A.~Tseytlin,
\textit{``{Superstring action in AdS(5) x S**5. Kappa symmetry light cone
  gauge}''},
\textsf{\doiref{10.1103/PhysRevD.63.046002}{Phys.Rev.~D63,~046002~(2001)}},
\texttt{\arxivref{hep-th/0007036}{hep-th/0007036}}.
%%CITATION = HEP-TH/0007036;%%

\bibitem{MTT2000}
R.~Metsaev, C.~B.~Thorn and A.~A.~Tseytlin,
\textit{``{Light cone superstring in AdS space-time}''},
\textsf{\doiref{10.1016/S0550-3213(00)00712-4}{Nucl.Phys.~B596,~151~(2001)}},
\texttt{\arxivref{hep-th/0009171}{hep-th/0009171}}.
%%CITATION = HEP-TH/0009171;%%

\bibitem{Gubser:2002tv}
S.~S.~Gubser, I.~R.~Klebanov and A.~M.~Polyakov,
\textit{``{A Semiclassical limit of the gauge / string correspondence}''},
\textsf{\doiref{10.1016/S0550-3213(02)00373-5}{Nucl.~Phys.~B636,~99~(2002)}},
\texttt{\arxivref{hep-th/0204051}{hep-th/0204051}}.
%%CITATION = HEP-TH/0204051;%%

\bibitem{Giombi}
S.~Giombi, R.~Ricci, R.~Roiban, A.~Tseytlin and C.~Vergu,
\textit{``{Quantum AdS(5) x S5 superstring in the AdS light-cone gauge}''},
\textsf{\doiref{10.1007/JHEP03(2010)003}{JHEP~1003,~003~(2010)}},
\texttt{\arxivref{0912.5105}{arxiv:0912.5105}}.
%%CITATION = ARXIV:0912.5105;%%

\bibitem{Giombi:2010bj}
S.~Giombi, R.~Ricci, R.~Roiban and A.~A.~Tseytlin,
\textit{``{Quantum dispersion relations for excitations of long folded spinning
  superstring in $AdS_5 \times S^5$}''},
\textsf{\doiref{10.1007/JHEP01(2011)128}{JHEP~1101,~128~(2011)}},
\texttt{\arxivref{1011.2755}{arxiv:1011.2755}}.
%%CITATION = ARXIV:1011.2755;%%

\bibitem{Giombi:2010zi}
S.~Giombi, R.~Ricci, R.~Roiban and A.~A.~Tseytlin,
\textit{``{Two-loop AdS5xS5 superstring: testing asymptotic Bethe ansatz and
  finite size corrections}''},
\textsf{\doiref{10.1088/1751-8113/44/4/045402}{J.~Phys.~A44,~045402~(2011)}},
\texttt{\arxivref{1010.4594}{arxiv:1010.4594}}.
%%CITATION = ARXIV:1010.4594;%%

\bibitem{Luscher:2008tw}
M.~Luscher and F.~Palombi,
\textit{``{Fluctuations and reweighting of the quark determinant on large
  lattices}''},
\textsf{\doiref{10.22323/1.066.0049}{PoS~LATTICE2008,~049~(2008)}},
\texttt{\arxivref{0810.0946}{arxiv:0810.0946}},
in: \textit{``{Proceedings, 26th International Symposium on Lattice field
  theory (Lattice 2008): Williamsburg, USA, July 14-19, 2008}''},
049p.
%%CITATION = ARXIV:0810.0946;%%

\bibitem{montvay}
I.~Montvay and G.~Muenster,
in: \textit{``Quantum Fields on a Lattice''},
Cambridge University Press.
%%CITATION = HEP-LAT/0306017;%%

\bibitem{Catterall:2015zua}
S.~Catterall,
\textit{``{Fermion mass without symmetry breaking}''},
\textsf{\doiref{10.1007/JHEP01(2016)121}{JHEP~1601,~121~(2016)}},
\texttt{\arxivref{1510.04153}{arxiv:1510.04153}}.
%%CITATION = ARXIV:1510.04153;%%

\bibitem{Catterall:2016dzf}
S.~Catterall and D.~Schaich,
\textit{``{Novel phases in strongly coupled four-fermion theories}''},
\texttt{\arxivref{1609.08541}{arxiv:1609.08541}}.
%%CITATION = ARXIV:1609.08541;%%

\bibitem{Finkenrath:2013soa}
J.~Finkenrath, F.~Knechtli and B.~Leder,
\textit{``{One flavor mass reweighting in lattice QCD}''},
\textsf{\doiref{10.1016/j.nuclphysb.2013.10.019,
  10.1016/j.nuclphysb.2014.01.019}{Nucl.~Phys.~B877,~441~(2013)}},
\texttt{\arxivref{1306.3962}{arxiv:1306.3962}},
[Erratum: Nucl. Phys.B880,574(2014)].
%%CITATION = ARXIV:1306.3962;%%

\bibitem{wimmer2012algorithm}
M.~Wimmer,
\textit{``Algorithm 923: Efficient numerical computation of the pfaffian for
  dense and banded skew-symmetric matrices''},
\textsf{ACM~Transactions~on~Mathematical~Software~(TOMS)~38,~30~(2012)}.

\bibitem{Wolff:2003sm}
ALPHA Collaboration, U.~Wolff,
\textit{``{Monte Carlo errors with less errors}''},
\textsf{\doiref{10.1016/S0010-4655(03)00467-3,
  10.1016/j.cpc.2006.12.001}{Comput.~Phys.~Commun.~156,~143~(2004)}},
\texttt{\arxivref{hep-lat/0306017}{hep-lat/0306017}},
[Erratum: Comput. Phys. Commun.176,383(2007)].
%%CITATION = HEP-LAT/0306017;%%

\bibitem{RHMC1}
A.~D.~Kennedy, I.~Horvath and S.~Sint,
\textit{``{A New exact method for dynamical fermion computations with nonlocal
  actions}''},
\textsf{\doiref{10.1016/S0920-5632(99)85217-7}{Nucl.~Phys.~Proc.~Suppl.~73,~834~(1999)}},
\texttt{\arxivref{hep-lat/9809092}{hep-lat/9809092}},
in: \textit{``{Lattice Field Theory. Proceedings: 16th International Symposium,
  Lattice '98, Boulder, USA, Jul 13-18, 1998}''},
834-836p.
%%CITATION = HEP-LAT/9809092;%%

\bibitem{RHMC2}
M.~A.~Clark and A.~D.~Kennedy,
\textit{``{The RHMC algorithm for two flavors of dynamical staggered
  fermions}''},
\textsf{\doiref{10.1016/S0920-5632(03)02732-4}{Nucl.~Phys.~Proc.~Suppl.~129,~850~(2004)}},
\texttt{\arxivref{hep-lat/0309084}{hep-lat/0309084}},
in: \textit{``{Lattice field theory. Proceedings, 21st International Symposium,
  Lattice 2003, Tsukuba, Japan, July 15-19, 2003}''},
850-852p,
[,850(2003)].
%%CITATION = HEP-LAT/0309084;%%

\bibitem{Basso:2010in}
B.~Basso,
\textit{``{Exciting the GKP string at any coupling}''},
\textsf{\doiref{10.1016/j.nuclphysb.2011.12.010}{Nucl.~Phys.~B857,~254~(2012)}},
\texttt{\arxivref{1010.5237}{arxiv:1010.5237}}.
%%CITATION = ARXIV:1010.5237;%%

\bibitem{Bruno:2014lra}
M.~Bruno, P.~Korcyl, T.~Korzec, S.~Lottini and S.~Schaefer,
\textit{``{On the extraction of spectral quantities with open boundary
  conditions}''},
\textsf{\doiref{10.22323/1.214.0089}{PoS~LATTICE2014,~089~(2014)}},
\texttt{\arxivref{1411.5207}{arxiv:1411.5207}},
in: \textit{``{Proceedings, 32nd International Symposium on Lattice Field
  Theory (Lattice 2014): Brookhaven, NY, USA, June 23-28, 2014}''},
089p.
%%CITATION = ARXIV:1411.5207;%%

\bibitem{Bruno:2014jqa}
M.~Bruno et~al.,
\textit{``{Simulation of QCD with N$_{f} =$ 2 $+$ 1 flavors of
  non-perturbatively improved Wilson fermions}''},
\textsf{\doiref{10.1007/JHEP02(2015)043}{JHEP~1502,~043~(2015)}},
\texttt{\arxivref{1411.3982}{arxiv:1411.3982}}.
%%CITATION = ARXIV:1411.3982;%%

\end{thebibliography}

\end{document}